\documentclass[aps,pre,10pt,showpacs,floatfix,onecolumn,nofootinbib,a4paper]{revtex4-2}
\usepackage{amssymb, amsmath, amsthm, mathtools}
\usepackage{bm}
\usepackage{mathrsfs}
\usepackage{hyperref,bookmark}
\usepackage{graphicx,colortbl}
\usepackage{epsfig,color}
\usepackage{mathrsfs}
\usepackage{verbatim}
\usepackage{url}
\usepackage{subcaption}
\usepackage{float}
\usepackage[greek,english]{babel}
\usepackage{dcolumn}
\usepackage{accents}

\usepackage{cases}

\usepackage{geometry}
\usepackage{paralist}

\usepackage{color}
\definecolor{lightgray}{gray}{0.8}

\newcommand{\tr}{\operatorname{tr}}
\newcommand{\dd}{\operatorname{d}\!}

\newcommand{\diver}{\operatorname{div}}
\newcommand{\curl}{\operatorname{curl}}

\newcommand{\n}{\bm{n}}
\newcommand{\e}{\bm{e}}

\newcommand{\normal}{\bm{\nu}}
\newcommand{\body}{\mathscr{B}}
\newcommand{\free}{\mathscr{F}}
\newcommand{\boundary}{\partial\mathscr{B}}
\newcommand{\Req}{R_\mathrm{e}}

\newcommand{\vt}{\vartheta}

\newcommand{\Q}{\mathbf{Q}}

\newcommand{\arctanh}{\operatorname{arctanh}}

\newcommand{\trans}{^\mathsf{T}}

\newcommand{\WOF}{W_\mathrm{OF}}
\newcommand{\WQT}{W_\mathrm{QT}}

\newcommand{\bend}{\bm{b}}

\newcommand{\Wn}{\mathbf{W}(\n)}
\newcommand{\Pn}{\mathbf{P}(\n)}
\newcommand{\Dn}{\mathbf{D}}
\newcommand{\I}{\mathbf{I}}
\newcommand{\twon}{(\n_1,\n_2)}
\newcommand{\zero}{\bm{0}}

\newcommand{\nT}{\n_{\mathrm{T}}}

\newcommand{\x}{\bm{x}}
\newcommand{\R}{\mathbf{R}}

\newcommand{\framec}{(\e_r,\e_\vt,\e_z)}
\newcommand{\region}{\mathscr{R}}

\newcommand{\grads}{\nabla_{\!\mathrm{s}}}
\newcommand{\temp}{\mathrm{T}}

\theoremstyle{definition}

\begin{document}
\latintext

\title{A Review on Phenomenological Models for Chromonic Liquid Crystals}
\author{Silvia Paparini}
	\email{silvia.paparini@unipd.it}	
\affiliation{Dipartimento di Matematica “Tullio Levi-Civita”, Universit`a degli Studi di Padova, Padua, Italy, and
Gruppo Nazionale per la Fisica Matematica, Istituto Nazionale di Alta Matematica “Francesco Severi”, Roma, Italy.}

\begin{abstract}
Chromonic liquid crystals (CLCs) are lyotropic materials which are attracting growing interest for their adaptability to living systems.
A considerable body of works has been devoted to exploring their properties and applications. In this paper, I endeavour to review some of the contributions concerning their theoretical modelling, aimed at rationalizing experimental observations. The intention is to present these developments within a unified framework, highlighting recent advances in the modelling of CLCs in both three-dimensional and two-dimensional geometries.\\
The elastic theory of CLCs is not completely established. Their ground state in the $3D$ space, as revealed by a number of recent experiments, is quite different from that of ordinary nematic liquid crystals: it is twisted instead of uniform. The common explanation provided for this state within the classical Oseen-Frank elastic theory demands that one Ericksen’s inequality is violated. 
Since such a violation would make the Oseen-Frank stored-energy density unbounded below, the legitimacy of these theoretical treatments is threatened by
a number of mathematical issues. To overcome these difficulties, a novel elastic theory has been proposed and tested for CLCs; it extends the classical Oseen–Frank energy by incorporating a quartic twist term.\\
Another key characteristic of CLCs is that they exhibit broad biphasic regions, in which the nematic and isotropic phases coexist. Mathematical models inspired by experimental settings have been developed for CLC droplets in two spatial dimensions. The contributions reviewed here address the morphogenesis of nuclei and topological defects during phase transitions, the topological shape transformations arising from the interplay of nematic elastic constants, and the prediction of shape bistability (yet to be observed) where tactoids (pointed, zeppelin-shaped droplets) and smooth-edged discoids can coexist in equilibrium. General methods have also been applied to experimental data to extract estimates of the isotropic surface tension at the nematic–isotropic interface and the chromonics' planar anchoring strength.\\
There are promising avenues for future research, including experimental validation of theoretical predictions and further theoretical challenges that remain to be addressed.
\end{abstract}

\date{\today}

\pagenumbering{arabic}

\maketitle

\section{Introduction}\label{sec:intro}
Chromonic liquid crystals (CLCs), also called \emph{chromonics}, form a very peculiar class of liquid crystals (LCs). These latter are anisotropic fluids that fall basically into two broad categories: they are either \emph{thermotropic} or \emph{lyotropic}, depending on whether it is temperature or concentration, respectively, responsible for driving the formation of these fascinating intermediate phases of soft condensed matter, which are birefringent like crystals and flow like liquids.

In particular, CLCs are lyotropic. They are composed of plank-like molecules with a poly-aromatic core and polar peripheral groups, aggregated in columnar stacks resulting from noncovalent attractions between the poly-aromatic cores. CLCs are formed by certain dyes, drugs, and short nucleic-acid oligomers in aqueous solutions \cite{dickinson:aggregate,tam-chang:chromonic,mariani:small,zanchetta:phase,nakata:end-to-end,fraccia:liquid}. Since most biological processes take function normally in these types of solutions, 
it is no wonder that interest in CLCs has recently surged for possible applications in medical sciences. 
But this is not the only reason that makes them special (or rather unique). A number of informative, updated reviews are available on this topic \cite{lydon:chromonic_1998,lydon:handbook,lydon:chromonic_2010,lydon:chromonic,dierking:novel}; they also witness the scientific interest surrounding this theme.

At low concentrations, supramolecular columns are not long enough to induce local nematic order. Upon increasing concentration, a nematic phase takes eventually over, although the critical concentration turns out to be much lower (see, for example, \cite{nayani:spontaneous}) than that predicted by Onsager's excluded volume theory \cite{onsager:effects}.  
Several unconventional models for columnar organization, envisioning the possibility that molecular stacks be either Y-shaped or side-slipped \cite{xiao:structural,park:self-assembly}, have been put forward to try and explain this discrepancy. Upon further increasing concentration, the nematic phase gives way to a phase, called the M phase, as it is similar (that is, it has a \emph{herringbone} texture similar) to the \emph{middle} phases of conventional amphiphile systems \cite{lydon:chromonic}.

CLCs are odd nematic phases. They do not seem to possess the same ground state as ordinary nematics. When a low-molecular liquid crystal in the nematic phase is left to itself, in the absence of either external disturbing agencies or confining boundaries, the nematic director $\n$, which represents on a macroscopic scale the average orientation of the elongated molecules that constitute the medium, tends to be uniform in space, in a randomly chosen direction. This is \emph{not} what a CLC does.

\begin{figure}[h]
	\centering
\begin{subfigure}[c]{0.45\linewidth}
	\centering
	\includegraphics[width=0.7\linewidth]{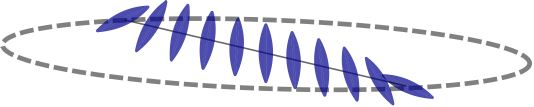}
\end{subfigure}
\quad
\begin{subfigure}{0.45\linewidth}
	\centering
	\includegraphics[width=0.7\linewidth]{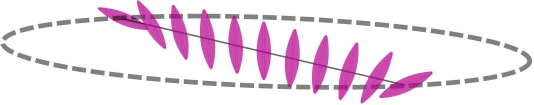}
\end{subfigure}
	\caption{Sketches representing the director arrangement in two symmetric variants of the escaped-twist (ET) distortion within a capillary tube with degenerate planar anchoring conditions on its lateral boundary (schematically indicated by a dashed circle). Figs. reprinted by \cite{paparini:elastic}.}
	\label{fig:ET_sketch}
\end{figure}

Experiments have been performed with these materials in capillary tubes, with either circular \cite{nayani:spontaneous,davidson:chiral} or rectangular \cite{fu:spontaneous} cross-sections, as well as on cylindrical shells \cite{javadi:cylindrical}, all enforcing \emph{degenerate planar} anchoring, which allows constituting columns to glide freely on the anchoring surface, provided they remain tangent to it. These experiments revealed that the spontaneous distortion is \emph{not} the alignment along the cylinder's axis, which is the only uniform one compatible with the boundary conditions; rather, it is a \emph{twisted} orientation \emph{escaped} along the cylinder's axis, swinging away from it on the cylinder's lateral boundary (see Fig.~\ref{fig:ET_sketch}).
Such an \emph{escaped-twist} (ET) distortion\footnote{Escaped-twist (ET) is a name perhaps first used in in \cite{ondris-crawford:curvature} for what before had been called \emph{twist-bend}  in \cite{cladis:non-singular} or \emph{escaped} in the third dimension in \cite{meyer:existence}.} is very similar (but not completely identical) to the \emph{double twist} (DT), which, when energetically  favoured in cholesteric liquid crystals, gives rise to their \emph{blue phases}, \cite{kleman:soft,meiboom:theory,meiboom:lattice,pisljar:blue}.

ET distortions come with two types of handiness: the director may wind either clockwise or anticlockwise as we progress radially outwards from the cylinder's axis. Being both helicities equally energetic, they are seen with equal probability, and so either singular (point) defects \cite{davidson:chiral} or regular domain walls \cite{nayani:spontaneous} may arise where two ET domains with opposite chirality come together. 

The elastic theory of chromonic liquid crystals is not completely established. Despite the lack of \emph{uniformity} in the ground state of these phases,\footnote{The classification of the most general \emph{uniform} distortions, which can fill the whole three-dimensional space, is given in \cite{virga:uniform} and recalled in Sect.~\ref{sec:ground}.} their curvature elasticity has been modeled by the Oseen-Frank theory, albeit with an anomalously small twist constant $K_{22}$. To accommodate the experimental findings and justify the twisted ground state, this constant has to be smaller than the saddle-splay constant $K_{24}$, in violation of one of the inequalities Ericksen~\cite{ericksen:inequalities} had put forward to guarantee that the Oseen-Frank stored energy be bounded below. 

After a brief summary of the classical theory for nematics to make the review self-contained, the issue as to whether ET distortions may or may not embody the ground state of chromonics is tackled in Sect.~\ref{sec:ground}. In this context, we also make contact with a notion of elastic frustration (Sec.\ref{sec:stability}), which naturally arises from the geometric incompatibility of ET distortions. Even though violating the Ericksen’s inequality does not prevent the Oseen-Frank stored-energy density to be well behaved in rigidly confined systems, applying the classical Oseen–Frank theory to free-boundary problems, such as those concerning the equilibrium shape of CLC droplets surrounded by their isotropic phase, can yield paradoxical results. Contrary to experimental evidence, they are predicted to
dissolve in a plethora of unstable smaller droplets. Therefore, both the observed ground state of CLCs and their ability to form stable twisted tactoids cannot be justified within the Oseen-Frank theory (Sec.~\ref{sec:conundrum}). To remedy this state of affairs,  a \emph{quartic} elastic theory was proposed for CLCs (Sec.~\ref{sec:quartic}) which alters the Oseen-Frank energy density by the addition of a \emph{single} quartic term in the \emph{twist} measure of nematic distortion. Preliminary experimental confirmations of the validity of this theory are recalled in Sec.~\ref{sec:twisted}. The spiralling texture of twisted hedgehogs in spherical cavities enforcing homeotropic anchoring is characterised by an inversion ring, which can be observed optically. Measurements of the inversion ring are contrasted with both the classical Oseen-Frank theory and the quartic twist theory, and shown to be in better accord with the
latter than with the former. The quartic theory features a phenomenological length $a$, whose measure is extracted from the data and shown to be fairly independent of the cavity radius, as expected for a material constant (Sec.~\ref{sec:estimates}).

Applications of CLCs in the life sciences rely on a proper characterization of these materials, including the determination of elastic constants, the isotropic
surface tension at the nematic/isotropic solution interface $\gamma$, and the anchoring strength on rigid substrates $\sigma_0$.
In this review, we focus primarily on the characterization of $\gamma$ and $\sigma_0$, while referring the reader to~\cite{varytimiadou:liquid} for a recent review on the characterization of elastic constants.
CLCs exhibit broad biphasic regions in which the nematic phase coexists with the isotropic phase. This feature is central to the final Section of this review (Sec.~\ref{sec:2dgeometries}), which is dedicated to $2d$ geometries in which bipolar CLC droplets in the nematic phase are surrounded by the isotropic phase and sandwiched between two parallel substrates. In this two-dimensional setting, we do not delve on the possibly controversial issue concerning the paradoxical consequences of violating the Erciknen's inequality, since neither $K_{22}$ nor $K_{24}$ play a role. After the introduction of a unified theoretical scene, we review key contributions that address the morphogenesis of nuclei and topological defects during phase transitions in CLCs, as well as topological shape transformation caused by the interplay of nematic elastic constants (Sec.~\ref{sec:morphogenesis}). 
Sec.~\ref{sec:shape} then discusses the theoretical prediction of a regime of shape coexistence, yet to be observed experimentally, which appears characteristic of the two-dimensional setting. In this regime, a range of droplet areas is identified where two distinct shapes could be observed, one tactoidal and the other discoidal (smooth), both bearing a bipolar arrangement of $\n$. Finally, Secs.~\ref{sec:estimategamma} and \ref{sec:estimatesigma} recall the methods used to estimate $\gamma$ and $\sigma_0$, respectively, from experimental data.

In Sec.~\ref{sec:conclusion}, we comment on promising avenues for future research, including the experimental validation of theoretical predictions and further theoretical challenges.

\section{Ground State}\label{sec:ground}
The contentious point to be addressed in this Section is whether the elasticity of chromonics can be adequately described by the Oseen-Frank’s classical theory of nematic liquid crystals. We begin by summarizing this theory and its formulation in a novel, equivalent way that serves better our purpose. Then, we address the problem posed by the peculiar ground state exhibited by CLCs.

The classical elastic theory of liquid crystals goes back to the pioneering works of Oseen~\cite{oseen:theory} and Frank~\cite{frank:theory}\footnote{Also a paper by Zocher~\cite{zocher:effect}, mainly concerned with the effect of a magnetic field on director distortions, is often mentioned among the founding contributions. Some go to the extent of also naming the theory after him. Others, in contrast, name the theory only after Frank, as they only deem his contribution to be fully aware of the nature of $\n$ as a \emph{mesoscopic}  descriptor of molecular order.}. This theory is variational in nature, as it is based on a bulk free energy functional $\free_\mathrm{b}$ written in the form
\begin{equation}
	\label{eq:free_energy}
	\free_\mathrm{b}[\n]:=\int_{\body}\WOF(\n,\nabla\n)\dd V,
\end{equation}
where $\body$ is a region in space occupied by the material and $V$ is the volume measure. In \eqref{eq:free_energy}, $\WOF$ measures the distortional cost produced by a deviation from a uniform director field $\n$. It is chosen to be the most general frame-indifferent,\footnote{A function $W(\n,\nabla\n)$ is \emph{frame-indifferent} if it is invariant under the action of the orthogonal group $\mathsf{O}(3)$, that is, if $W(\Q\n,\Q(\nabla\n)\Q\trans)=W(\n,\nabla\n)$ for all $\Q\in\mathsf{O}(3)$, where $\Q\trans$ denotes the transpose of $\Q$.}  even function quadratic in $\nabla\n$, 
\begin{equation}
	\label{eq:free_energy_density}
	\WOF(\n,\nabla\n):=\frac{1}{2}K_{11}\left(\diver\n\right)^2+\frac{1}{2}K_{22}\left(\n\cdot\curl\n\right)^2+ \frac{1}{2}K_{33}|\n\times\curl\n|^{2} + K_{24}\left[\tr(\nabla\n)^{2}-(\diver\n)^{2}\right].
\end{equation}
Here $K_{11}$, $K_{22}$, $K_{33}$, and $K_{24}$ are elastic constants characteristic of the material. They are often referred to as the \emph{splay}, \emph{twist}, \emph{bend}, and \emph{saddle-splay} constants, respectively, by the features of the different orientation fields, each with a distortion energy proportional to a single term in \eqref{eq:free_energy_density} (see, for example, Ch.~3 of \cite{virga:variational}). 

Recently, Selinger~\cite{selinger:interpretation} has reinterpreted the classical formula \eqref{eq:free_energy_density} by decomposing the saddle-splay mode into a set of other independent modes. The starting point of this decomposition is a novel representation of $\nabla\n$ (see also \cite{machon:umbilic}),
\begin{equation}
	\label{eq:nabla_n_novel}
	\nabla\n=-\bend\otimes\n+\frac12T\Wn+\frac12S\Pn+\Dn,
\end{equation}
where $\bend:=-(\nabla\n)\n=\n\times\curl\n$ is the \emph{bend} vector, $T:=\n\cdot\curl\n$ is the \emph{twist}, $S:=\diver\n$ is the \emph{splay}, $\Wn$ is the skew-symmetric tensor that has $\n$ as axial vector, $\Pn:=\I-\n\otimes\n$ is the projection onto the plane orthogonal to $\n$, and $\Dn$ is a symmetric tensor such that $\Dn\n=\zero$ and $\tr\Dn=0$. By its own definition, $\Dn\neq\zero$ admits the following biaxial representation,
\begin{equation}
	\label{eq:D_representation}
	\Dn=q(\n_1\otimes\n_1-\n_2\otimes\n_2),
\end{equation}
where $q>0$ and $\twon$ is a pair of orthogonal unit vectors in the plane orthogonal to $\n$, oriented so that $\n=\n_1\times\n_2$.\footnote{It is argued in \cite{selinger:director} that $q$ should be given the name \emph{tetrahedral} splay, to which we would actually prefer \emph{octupolar} splay for the role played by a cubic (octupolar) potential on the unit sphere \cite{pedrini:liquid} in representing all scalar measures of distortion,  but $T$.}

By use of the following identity, 
\begin{equation}
	\label{eq:identity}
	2q^2=\tr(\nabla\n)^2+\frac12T^2-\frac12S^2,
\end{equation}
we can easily give \eqref{eq:free_energy_density} the equivalent form
\begin{equation}
	\label{eq:Frank_equivalent}
	\WOF(\n,\nabla\n)=\frac12(K_{11}-K_{24})S^2+\frac12(K_{22}-K_{24})T^2+\frac12K_{33}B^2+2K_{24}q^2,
\end{equation}
where $B^2:=\bend\cdot\bend$. Since $(S,T,B,q)$ are all independent \emph{distortion characteristics}, it readily follows from \eqref{eq:Frank_equivalent} that $\WOF$ is positive semi-definite whenever
\begin{subequations}\label{eq:Ericksen_inequalities}
\begin{eqnarray}
	K_{11}&\geqq& K_{24}\geqq0,\label{eq:Ericksen_inequalities_1}\\
	K_{22}&\geqq& K_{24}\geqq0, \label{eq:Ericksen_inequalities_2}\\
	K_{33}&\geqq&0,\label{eq:Ericksen_inequalities_3}
\end{eqnarray}
\end{subequations}
which are the celebrated \emph{Ericksen's inequalities} \cite{ericksen:inequalities}.

If these inequalities are satisfied in strict form, the global ground state of $\WOF$ is attained on the uniform director field, characterized by
\begin{equation}
	\label{eq:uniform_ground_state}
	S=T=B=q=0.
\end{equation}

More generally, it has been shown \cite{virga:uniform} that besides \eqref{eq:uniform_ground_state} the only \emph{uniform distortions}, that is, director fields that fill three-dimensional Euclidean space, having everywhere the same distortion characteristics, are only those for which
\begin{equation}
	\label{eq:uniform_distortions}
	S=0,\quad T=\pm2q,\quad b_1=\pm b_2=b,
\end{equation}
corresponding to Meyers's \emph{heliconical} distortions \cite{meyer:structural} characterizing the ground state of the \emph{twist-bend}
nematic phases identified  experimentally in \cite{cestari:phase}\footnote{In \eqref{eq:uniform_distortions}, $q$ is \emph{positive} and $b$ arbitrary. As shown in \cite{virga:uniform}, if $q$ vanishes also does $b$ and both forms of uniform distortions reduce to the standard uniform orientation in \eqref{eq:uniform_ground_state}.}.

The experimental evidence gathered in \cite{davidson:chiral,nayani:spontaneous} tells us that CLCs within a cylinder with degenerate planar anchoring on the lateral wall acquire either of ET distortions (see Fig.~\ref{fig:ET_sketch}). This shows that the uniform distortion in  \eqref{eq:uniform_ground_state} is \emph{not} the ground state of chromonics, and neither are \eqref{eq:uniform_distortions}, as we shall see, thus entailing a degree of \emph{elastic frustration} in the ground state.

\subsection{ET configurations in a circular cylinder}

ET distortions were first described analytically by Burylov~\cite{burylov:equilibrium}\footnote{The same results arrived at in \cite{burylov:equilibrium} were independently reobtained in \cite{davidson:chiral}. See also \cite{paparini:stability} for detailed computations.} ; they exist as solutions to the pertinent Euler-Lagrange equations only when the uniform orientation along the cylinder's axis ceases to be locally stable. We elaborate on this in this Section.
For $\body$ a circular cylinder, we now describe the ET distortion that minimizes the free-energy functional $\free_\mathrm{b}$ in \eqref{eq:free_energy} subject to the \emph{planar degenerate} anchoring, for which, 
\begin{equation}\label{eq:planar_degenerate}
	\n\cdot\normal\equiv0\quad\text{on}\quad\boundary.
\end{equation}

Here, the geometry is rigid and, under the assumption \eqref{eq:planar_degenerate}, the additional surface energy 
can be treated as an inessential additive constant. This will not be the case in the following Section, where the region $\body$ is no longer fixed. 

Let $R$ be the radius of the cylinder $\body$ and $L$ its height. We assume that in the frame $\framec$ of cylindrical coordinates $(r,\vt,z)$, with $\e_z$ along the axis of $\body$, $\n$ is represented as
\begin{equation}
	\label{eq:n_bur}
	\n=\sin\beta(r)\e_\vt+\cos\beta(r)\e_z,
\end{equation}
where the \emph{polar} angle $\beta\in[-\pi,\pi]$, which $\n$ makes with the cylinder's axis, depends only on the radial coordinate $r$. 

By changing the variable $r$ into
\begin{equation}\label{eq:rho_definition}
	\rho:=\frac{r}{R},
\end{equation}
which ranges in $[0,1]$, we arrive at the following reduced functional, $\mathcal{F}[\beta]$, which is an appropriate dimensionless form of Frank's free-energy functional $\free_\mathrm{b}$,
\begin{equation}
	\label{eq:free_bur}
	\mathcal{F}[\beta]:=\frac{\free_\mathrm{b}[\n]}{2\pi K_{22}L}= \int_0^1\left(\frac{\rho\beta'^2}{2}+\dfrac{1}{2\rho}\cos^2\beta\sin^2\beta+\dfrac{k_3}{2\rho}\sin^4\beta\right) \dd\rho+\frac12(1-2k_{24})\sin^2\beta(1),
\end{equation}
where 
the following scaled elastic constants have been introduced,
\begin{equation}
	\label{eq:scaled_elastic_constant}
	k_3 :=\frac{K_{33}}{K_{22}}>0, \quad k_{24}:=\frac{K_{24}}{K_{22}}>0.
\end{equation}
For the integral in \eqref{eq:free_bur} to be convergent, $\beta$ must be subject to the condition 
\begin{equation}
	\label{eq:boundary_condition_0}
	\beta(0)=0,
\end{equation}
which amounts to require that $\n$ is along $\e_z$ on the cylinder's axis.\footnote{Actually, the convergence requirement would also be satisfied by enforcing the more general condition $\sin\beta(0)=0$; however, this choice is not restrictive, it rather rests on the nematic symmetry, for which $\n$ and $-\n$ are physically equivalent.} The functions $\beta=\beta(\rho)$ of class $\mathcal{C}^2$ on $[0,1]$ that satisfy \eqref{eq:boundary_condition_0} and make \eqref{eq:free_bur} stationary, are given by
\begin{equation}
	\label{eq:bur_solution}
	 \beta_\mathrm{ET}(\rho):=\arctan\left(\frac{2\sqrt{k_{24}(k_{24}-1)}\rho}{\sqrt{k_{3}}\left[k_{24}-(k_{24}-1)\rho^2\right]}\right), 
\end{equation}
and its opposite $- \beta_\mathrm{ET}$. They are permitted only if $k_{24}>1$, that is,  by \eqref{eq:scaled_elastic_constant}, if
\begin{equation}
\label{eq:inequalityviolated}
K_{24}>K_{22},
\end{equation}
and thus only if Ericksen's inequality \eqref{eq:Ericksen_inequalities_2} is violated. The solutions $\beta_\mathrm{ET}$ and $-\beta_\mathrm{ET}$ represent the two variants of the ET distortion (with opposite chiralities). Moreover, $\mathcal{F}[\beta_\mathrm{ET}]$ can be written explicitly in terms of the reduced elastic constants only as
\begin{equation}
	\label{eq:ET_free_energy}
	\mathcal{F}_\mathrm{ET}=
	\begin{cases}
		1-k_{24}+\frac{1}{2}\frac{k_3}{\sqrt{1-k_3}}\arctanh\left(\frac{2\sqrt{1-k_3}(k_{24}-1)}{k_3+2(k_{24}-1)}\right), \quad&k_3\leqq1,\\ 1-k_{24}+\frac{1}{2}\frac{k_3}{\sqrt{k_3-1}}\arctan\left(\frac{2\sqrt{k_3-1}(k_{24}-1)}{k_3+2(k_{24}-1)}\right),\quad&k_3\geqq1,
	\end{cases}
\end{equation}
an expression that, for $k_{24}>1$, can be shown to be negative in both instances, as the following inequalities hold true,
\begin{align}
	\label{eq:burylov_energy_inequalities}
1-k_{24}\leqq\mathcal{F}_\mathrm{ET}\leqq-\frac{2(k_{24}-1)^2}{2k_{24}-1}&<0,\quad\text{for}\quad0\leqq k_3\leqq1,\\
-\frac{2(k_{24}-1)^2}{2k_{24}-1}\leqq\mathcal{F}_\mathrm{ET}&<0,\quad\text{for}\quad k_3\geqq1.
\end{align}
$\mathcal{F}_\mathrm{ET}$, as given by \eqref{eq:ET_free_energy} as a function of $k_3$, is continuous along with its derivatives at $k_3=1$\footnote{Apart from a different scaling of the constant $K_{24}$, the formula in \eqref{eq:ET_free_energy} for $k_3\geqq1$ coincides with equation (5) of \cite{davidson:chiral}.}. Moreover, the same formula is also valid for the energy of the mirror image $-\beta_\mathrm{ET}$ of $\beta_\mathrm{ET}$. Thus, whenever the ET distortion is permitted, that is, for $k_{24}>1$ (and so for \eqref{eq:inequalityviolated}), it possesses less elastic free energy than the uniform alignment $\n\equiv\e_z$, and so it becomes eligible for the ground state of CLCs, at least within the cylindrical confinement investigated experimentally. Therefore, the price to pay to model mathematically the experimental observations with the ET distortion is to renounce one of Ericksen's inequalities, thus accepting that Frank's functional in \eqref{eq:free_energy} may be unbounded below, jeopardizing in general its coercivity. 

Chromonic liquid crystals include Sunset Yellow (SSY), a popular dye in food industry, and disodium cromoglycate (DSCG), an antiasthmatic drug. For both materials, the inequality in eq. \ref{eq:Ericksen_inequalities_2} is allegedly violated. For example, for a solution of SSY in water with concentration $c=30.0\%$ (wt/wt) at temperature $25\,^\circ\mathrm{C}$ the following values of the elastic constants were measured in \cite{zhou:elasticity_2012}, $K_{11} = 4.3\mathrm{pN}$, $K_{22} = 0.7\mathrm{pN}$, and $K_{33} = 6.1\mathrm{pN}$. At equilibrium, by measuring the polar angle $\beta$ at different distances from the capillary's axis, it was found in \cite{davidson:chiral} that $K_{24} = 15.8\mathrm{pN}>K_{22}$. Similarly, using the experimental data for DSCG available from \cite{zhou:elasticity_2012}, \cite{eun:effects} refers to an aqueous solution with concentration $c=14.0\%$ (wt/wt) at temperature $21.5\,^\circ\mathrm{C}$, for which $K_{33}/K_{22}=30$; by estimating the twist angle at the capillary wall through \eqref{eq:bur_solution} at $\rho=1$, they obtain $K_{24}/K_{22}=15>1$.

\subsection{Stability against the odds in rigidly confined systems}\label{sec:stability}
This Section is dedicated to illustrate why the violation of the Ericksen's inequality \eqref{eq:Ericksen_inequalities_2} in rigidly confined systems does not prevent the integral in \eqref{eq:free_energy} to be well-behaved, provided that the constants $K_{11}$, $K_{22}$, and $K_{33}$ are all positive. First, we shall make contact with a notion of elastic frustration, which arises naturally from the geometric incompatibility of ET distortions. This becomes evident when considering the distortion for which all characteristics vanish, but $T$.
It is \emph{not} uniform and cannot fill space; it can possibly be realized locally, but not everywhere. In words, we say that it is a \emph{frustrated} ground state. It is, however, relevant to CLCs that the double twist is attained exactly on the symmetry axis of both chiral variants of the ET field described by $\pm\beta_\mathrm{ET}$, where the boundary conditions has the least influence \cite{paparini:stability},
\begin{equation}
	\label{eq:distortion_measure_Burylov_rho_0}
	S=q=b_1=b_2=0, \quad T=\frac{4\sqrt{k_{24}-1}}{R\sqrt{k_3k_{24}}}.
\end{equation}
The length scale of the elastic frustration induced by cylindrical confinement appears explicitly  in \eqref{eq:distortion_measure_Burylov_rho_0} through the radius $R$.

In \cite{paparini:stability}, a central role in taming the unboundedness of the energy is played by the specific boundary conditions imposed in the experiments on the boundary of rigid containers, \eqref{eq:planar_degenerate}.
Instances are known in the literature where appropriate boundary conditions salvage a functional that in other, more general circumstances would fail to attain its minimum (see, for example, \cite{day:sphere}). Here a similar situation arises with the complicity of cylindrical symmetry. In this case, indeed, as remarked in \cite{koning:saddle-splay}, the $K_{24}$-integral can be rewritten as 
\begin{equation}\label{eq:K_24_geometric}
	-K_{24}\int_{\boundary}\left(\kappa_1n_1^2+\kappa_2n_2^2\right)\dd A\,,
\end{equation}
where $\kappa_1$ and $\kappa_2$ are the principal curvatures of $\boundary$, and $n_i$ are the components of $\n$ along the corresponding principal directions of curvature\footnote{We write the curvature tensor as $\grads\normal=\kappa_1\e_1\otimes\e_1+\kappa_2\e_2\otimes\e_2$, where $\e_1$ and $\e_2$ are unit vectors along the principal directions of curvature of $\boundary$.}. It is clear from \eqref{eq:K_24_geometric} that for $K_{24}>0$, which is the strong form of  \eqref{eq:Ericksen_inequalities}, whenever  \eqref{eq:planar_degenerate} applies the saddle-splay energy would locally tend to orient $\n$ on $\boundary$ along the direction of \emph{maximum} (signed) curvature. For a region $\body$ whose boundary $\boundary$ has bounded principal curvatures, the $K_{24}$-energy is then always finite, and so the following \emph{reduced} Ericksen's inequalities suffice to guarantee that $\free_\mathrm{b}$ is bounded below,
\begin{equation}
	\label{eq:Ericksen_ineqaulities_reduced}
	K_{11}\geqq0,\quad K_{22}\geqq0,\quad K_{33}\geqq0.
\end{equation}

The same reassuring conclusion was reached in \cite{long:violation}, which proposed that the pure (double) twist mode that would characterize the ground state of chromonics as a consequence of \eqref{eq:inequalityviolated}, being non-uniform and so unable to fill space, prompts the excitation of other elastic modes whose positive cost counterbalances the divergence to negative infinity of the total free energy.

The issue about the stability of such a solution then remained open, a question that was not idle to ask, given the wildness of the parent energy. To resolve this issue, \cite{paparini:stability} derives a general formula for the second variation of Frank's elastic free-energy functional and applied it to the study of the (local) stability of the twisted ground state of chromonics. It is concluded that this is stable, despite the violation of one
Ericksen’s inequality\footnote{A violation nonetheless necessary for these fields to be equilibrium solutions.}. The local stability of the ET distortions nurtures the hope that this violation does not pose a serious threat to the applicability of the Oseen-Frank's elastic theory to chromonics. But the question remained as to whether different boundary conditions, still physically significant, could unleash the unboundedness of the total free energy potentially related to the violation of one Ericksen inequality (see also \cite{long:violation} in this connection).

\subsection{Free-boundary paradoxes for CLCs}\label{sec:paradoxes}
Liquid crystals are (within good approximation) incompressible fluids. Thus, when the region $\body$ is \emph{not} fixed, as in the cases considered in this Section, for a given amount of material, $\body$ is subject to
the  \emph{isoperimetric} constraint that prescribes its volume,
\begin{equation}
	\label{eq:isoperimetric_constraint}
	V(\body)=V_0.
\end{equation}
When $\body$ is surrounded by an isotropic fluid, a surface energy arises at the free interface $\boundary$, which, following \cite{rapini:distortion}, we represent as
\begin{equation}
	\label{eq:surface_energy}
	\free_\mathrm{s}[\body;\n]:=\int_{\boundary}\gamma[1+\omega(\n\cdot\normal)^2]\dd A,
\end{equation}
where $\normal$ is the outer unit normal to $\boundary$, $\gamma>0$ is the \emph{isotropic} surface tension, and $\omega>-1$ is a dimensionless parameter weighting the \emph{anisotropic} component of surface tension. For $\omega>0$, $\free_\mathrm{s}$ promotes the \emph{degenerate planar}  anchoring, whereas for $\omega<0$, it promotes the \emph{homeotropic} anchoring. For free-boundary problems, the total free energy functional will then be written as
\begin{eqnarray}\label{eq:free_energy_total}
	\free_{\mathrm{t}}[\body;\n]:=\free_\mathrm{b}[\body;\n]+\free_\mathrm{s}[\body;\n],
\end{eqnarray}
and the domain $\body$, subject to \eqref{eq:isoperimetric_constraint}, is also an unknown to be determined so as to minimize $\free$. When the geometry is rigid as in Sec.~\ref{sec:stability}, under the assumption \eqref{eq:planar_degenerate}, the surface energy $\free_{\mathrm{s}}$ in \eqref{eq:surface_energy} can be treated as an inessential additive constant.

The question is now whether in this case the surface energy plays the stabilizing role of the boundary conditions in rigid containers and prevent the degeneration of the Frank's free energy to $-\infty$.
In \cite{paparini:paradoxes}, this question is answered for the negative. It is found that if $K_{22}<K_{24}$, a CLC droplet, tactoidal\footnote{\emph{Tactoids} are elongated, cylindrically symmetric shapes with pointed ends as poles.} in shape and surrounded by an isotropic fluid environment enforcing degenerate planar anchoring for the director, is predicted to be unstable against \emph{shape change}: it would split indefinitely in smaller tactoids while the total free energy plummets to negative infinity.  

\begin{figure}[h]
	\centering
\begin{subfigure}[c]{0.45\linewidth}
	\centering
	\includegraphics[width=1.1\linewidth]{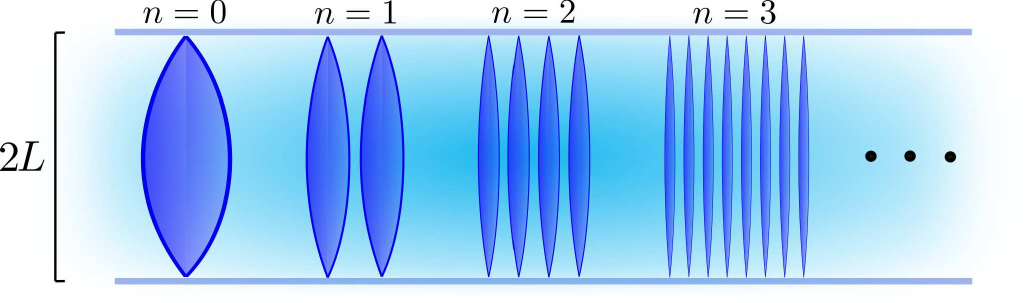}
	\caption{Each droplet splits in halves at every step, thus preserving the total volume. All drops have one and the same polar span $2L$.}
	\label{fig:crumblingparadox}
\end{subfigure}
$\qquad\ $
	\begin{subfigure}{0.45\linewidth}
	\centering
	\includegraphics[width=0.17\linewidth]{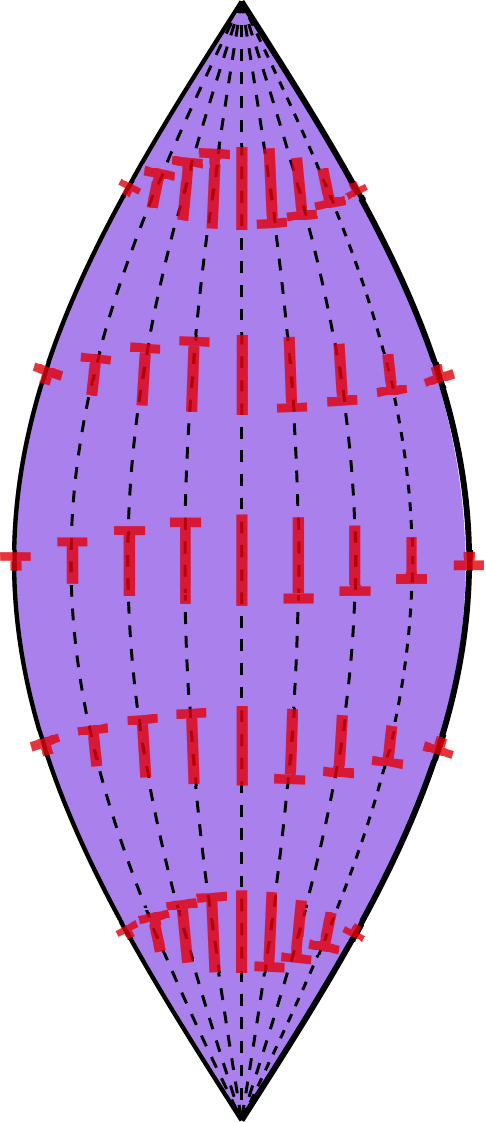}
	\caption{A tactoid with a twisted nematic director field represented as in \eqref{eq:n_bur_degenerate} and \eqref{eq:angle_alpha}.}
	\label{fig:tactoid}
	\end{subfigure}
\caption{Splitting procedure described in the text. The $(r,z)$ plane of the drawings is a symmetry plane of the drops through their axis. In Fig.~\ref{fig:tactoid}, a (red) segment represents $\n$ when it lies on the plane of the drawing, while a nail is used for the projection of $\n$ on that plane when the director is askew with it, the head designating conventionally the end on the same side as the viewer. Figs. reprinted by \cite{paparini:paradoxes}.}
\label{fig:tot_splitting}
\end{figure}

As an example of minimizing sequences with diverging energy which generates paradoxes within the Oseen-Frank theory, I report here the case of tactoidal drops confined between two parallel plates, $2L$ apart (see Fig.~\ref{fig:tot_splitting}).
Each drop occupies a region $\body$ in three-dimensional space rotationally symmetric about the $z$-axis of a standard cylindrical frame $(\e_r, \e_\vartheta, \e_z)$ whose boundary $\boundary$ is obtained by rotating the graph of a given smooth function, $R=R(z)$, which represents the radius of the drop's cross-section at height $z$. The sequence starts with a single parent drop of given volume $V_0$ with twisted director field represented by
\begin{equation}
\label{eq:n_bur_degenerate}
\n=\cos\alpha(z)\sin\beta_{\mathrm{ET}}\e_r+\sin\alpha(z)\sin\beta_{\mathrm{ET}}\e_\vartheta+\cos\beta_{\mathrm{ET}}\e_z,
\end{equation}
where $\alpha\in[0,2\pi)$ is the \emph{azimuthal} angle defined by
\begin{equation}
\label{eq:angle_alpha}
\alpha(z)=\begin{cases}
\arccos\left(\dfrac{R'(z)}{\tan\beta(1)}\right),\\\\
2\pi-\arccos\left(\dfrac{R'(z)}{\tan\beta(1)}\right), 
\end{cases}
\end{equation}
and $\beta_{\mathrm{ET}}\in[0,\pi]$ is the \emph{polar} angle and corresponds to the ET configuration in \eqref{eq:bur_solution} with now $\rho=r/R(z)\in[0,1]$. A constraint arises from \eqref{eq:angle_alpha} for $R'$, that is,
\begin{equation}
\label{eq:Rprime_constraint}
 -|\tan\beta(1)|\leqq R'(z)\leqq |\tan\beta(1)|,
\end{equation}
meaning that the drops have pointed tips. The director in \eqref{eq:n_bur_degenerate} is tangent to $\boundary$ and so fulfill the degenerate planar condition \eqref{eq:planar_degenerate}. Figure~\ref{fig:tactoid} illustrates our construction for a \emph{twisted tactoid}: it shows a meridian cross-section of the drop. 
It can be proved that $2L$ corresponds to the polar extension reached by all the drops.
 
Splitting recursively the parent drop in halves, preserving the total volume, as in Fig.~\ref{fig:crumblingparadox}, drive the total free energy to negative infinity. Indeed, by proceeding in steps indexed by the integer $n\in\mathbb{N}$, the total free energy $\free_{n}$ at the step $n$ satisfies the following estimate,
\begin{equation}
	\label{eq:F_n_estimate}
	\free_n\leqq2^nL\mathcal{F}_\mathrm{ET}+\sqrt{\frac{8}{3}}2^{n/2}\gamma \Req^{3/2} L^{1/2}+\mathcal{O}(2^{-n/2}),\quad n\to\infty,
\end{equation}  
where $\Req$ is the equivalent radius of the sphere of volume $V_0$, and $\mathcal{F}_\mathrm{ET}$ has the same value \eqref{eq:ET_free_energy}. Since $\mathcal{F}_\mathrm{ET}<0$ whenever $K_{24}>K_{22}$, \eqref{eq:F_n_estimate} implies the divergence of $\free_n$ to negative infinity as the splitting proceeds indefinitely. This confirms that the total free energy of a confined CLC drop is unbounded\footnote{As proved in \cite{paparini:paradoxes}, if the parent drop is splitted in appropriate unequal components, the CLC drop is nonetheless unstable against domain splitting.}. 

Other perplexing consequences of \eqref{eq:inequalityviolated} are also empathized in \cite{long:violation}.  These may involve the sensitivity of the material to the geometry of the container and its compatibility with impurities, such as dust or other colloidal particles.

\subsection{Conundrum}\label{sec:conundrum}
\begin{figure}
	\centering 
	\includegraphics[width=.33\linewidth]{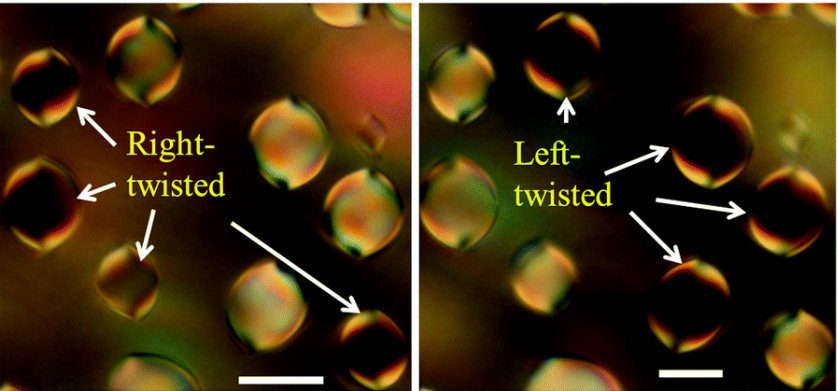}
	\caption{Textures of left- and right-twisted tactoids stable in the biphasic region. Fig. reprinted by \cite{peng:chirality}.}
	\label{fig:tactoid_observed}
\end{figure}

As recalled in Sec.~\ref{sec:stability}, violation of the Ericksen's inequality $K_{22}\geqq K_{24}$ \eqref{eq:Ericksen_inequalities_2} in the presence of degenerate planar anchoring is \emph{not} prejudicial to the stability of the twisted ground state; this has perhaps nurtured the hope that this inequality may  be renounced in the Oseen-Frank theory of CLCs. Sec.~\ref{sec:paradoxes} shows that this is not the case, as such a relaxed theory would entail shape instability of tactoids. This instability is in sharp contrast with the wealth of experimental observations of CLC tactoidal droplets, stable in the biphasic region of phase space, where nematic and isotropic phases coexist in equilibrium. Experiments have been carried out with a number of substances (including DSCG and SSY) stabilized by the addition of neutral (achiral) condensing agents (such as PEG and Spm) \cite{tortora:self-assembly,tortora:chiral,peng:chirality,nayani:using,shadpour:amplification}. These studies have consistently reported stable twisted bipolar tactoids (see, for example, Fig.~\ref{fig:tactoid_observed}).
Therefore, we cannot justify within the Oseen-Frank theory both the observed ground state of CLCs and their ability to form stable twisted tactoids. Two ways to avoid this contradiction can be seen: either 
\begin{inparaenum}[(i)]
	\item\label{item:case_one} the common interpretation of the capillary experiments that established the CLC ground state is incorrect, 
or	\item \label{item:case_two}  the Oseen-Frank theory is inapt to describe the elasticity of these materials. 
\end{inparaenum}

Boundary conditions are instrumental to alternative  \eqref{item:case_one}: one might question whether a mild azimuthal anchoring is at work, which could alter the determination of $K_{24}$ so that the Ericksen inequalities are not violated. It can be trusted that the detailed experimental analysis of the capillaries' inner boundary performed in \cite{davidson:chiral}  with the aid of both atomic force microscopy (AFM) and scanning electron microscopy (SEM);\footnote{See also the  supplementary information of \cite{davidson:chiral} and the AFM measurements of \cite{nayani:spontaneous}.} it was concluded that any azimuthal anchoring, if at all present, must be negligible compared with the saddle-splay energy, thus fully supporting the hypothesis of a pure  degenerate planar anchoring.

In want of further experimental data, recent studies are inclined towards alternative \eqref{item:case_two}: it is reckon worth pursuing a novel elastic theory for CLCs.
Some proposals have been advanced. For example, in \cite{long:violation} the role of added disclinations is advocated (provided that their energy cost can be made sufficiently low), whereas in \cite{paparini:elastic} a quartic twist term is added to the Oseen-Frank free energy density, which has the potential to restore shape stability when the Ericksen's inequality is violated. The following Section reviews the quartic twist approach, recently developed and tested in \cite{paparini:spiralling,ciuchi:inversion,paparini:what}.

\section{Quartic Twist Energy}\label{sec:quartic}
To overcome the difficulties arising from applying the Oseen-Frank theory to the curvature elasticity for chromonics, an elastic \emph{quartic twist theory} has recently been proposed in \cite{paparini:elastic}. The essential feature of this theory is to envision a double twist with two equivalent chiral variants as ground state of CLCs in three-dimensional space,
\begin{equation}
\label{eq:double_twist}
S = 0, \quad T = \pm T_0, \quad B = 0, \quad q = 0.
\end{equation}
The degeneracy of the ground  double twist  in \eqref{eq:double_twist} arises from the achiral nature of the molecular aggregates that constitute these materials, which is reflected in the lack of chirality of their condensed phases.

The elastic stored energy must equally penalize both ground chiral variants. The minimalistic proposal to achieve this goal is to add  a \emph{quartic twist} term to the Oseen-Frank stored-energy density:
\begin{equation}
\label{eq:quartic_free_energy_density}
\WQT(\n,\nabla\n)=\frac{1}{2}(K_{11}-K_{24})S^2+\frac{1}{2}(K_{22}-K_{24})T^2+ \frac{1}{2}K_{23}B^{2}+\frac{1}{2}K_{24}(2q)^2 + \frac14K_{22}a^2T^4,
\end{equation}
where $a$ is a \emph{characteristic length}. Unlike $\WOF$ when \eqref{eq:Ericksen_inequalities_2} is violated, $\WQT$ is bounded below whenever 
\begin{subequations}\label{eq:new_inequalities}
\begin{eqnarray}
	K_{11}&\geqq&K_{24}\geqq0,\label{eq:new_inequalities_1}\\
K_{24}&\geqq&K_{22}\geqq0, \label{eq:new_inequalities_2}\\
	K_{33}&\geqq&0.\label{eq:new_inequalities_3}
\end{eqnarray}
\end{subequations}
If these inequalities  hold, as we shall assume here, then $\WQT$ is minimum at the degenerate double-twist \eqref{eq:double_twist}
characterized by
\begin{equation}
	\label{eq:T_0min}
	T_0:=\frac{1}{a}\sqrt{\frac{K_{24}-K_{22}}{K_{22}}}.
\end{equation}
The parameter $a$ encodes the length scale over which distortions would be locally stored in the ground state. As to the physical size of such a length scale, it may be comprised in a wide range. While at the lower end we may place the persistence length of the molecular order, which characterizes the flexibility of CLC aggregates,\footnote{The persistence length of self-assembled flexible aggregates is a length over which unit vectors tangential to the aggregates lose correlation. For CLCs, it is estimated on the order of tens to hundreds of $\mathrm{nm}$ \cite{zhou:lyotropic}} the upper end is hard to make definite. It can be expected that $a$ would be exposed to the same indeterminacy that affects many (if not all) supramolecular structures in lyotropic systems. The most telling example is perhaps given by cholestric liquid crystals, which give rise to a chiral structure (characterized by a single twist $T=\pm2q$) starting from chiral molecules. If the macroscopic pitch (measured by $|1/T|$) were determined by the molecular chirality, it would result several orders of magnitude smaller than the observed ones. Here, $a$ will be treated as a phenomenological parameter, to be determined experimentally.

From now on, in the definition \eqref{eq:free_energy} for $\free_{\mathrm{b}}$ we shall replace $\WOF$ with $\WQT$. The quartic theory is built with the intent of curing the paradoxes encountered within the Oseen-Frank theory when handling free-boundary problems for chromonics. As recalled in Sec.~\ref{sec:paradoxes}, $\free_{\mathrm{t}}$ in \eqref{eq:free_energy_total} is proved to be \emph{unbounded} below in a class of free-boundary problems, if $\WOF$ is chosen to be the stored-energy density with \eqref{eq:Ericksen_inequalities_2} violated. Instead, the predicted disruptive mechanism cannot be at work within the quartic twist theory; as a consequence of the ground state of CLCs in $3D$ space envisioned by the quatic theory, $\free_{\mathrm{t}}$ obeys
\begin{equation}
	\label{eq:bound_F_t}
	\free_{\mathrm{t}}[\body;\n]\geqq -\frac{(K_{24}-K_{22})^2}{K_{22}}\frac{V_0}{4a^2}+\gamma_\omega(36\pi V_0^2)^{1/3},
\end{equation}
with
\begin{equation}
	\label{eq:gamma_omega}
	\gamma_\omega:=\gamma(1+\min\{0,\omega\})>0,
\end{equation}
thereby establishing a lower bound for $\free_{\mathrm{t}}$ that only depends on material constants and the prescribed volume $V_0$.

\subsection{Twisted Hedgehogs}\label{sec:twisted}
\begin{figure}[h]
	\centering
\begin{subfigure}{0.45\linewidth}
	\centering
	\includegraphics[width=0.783\linewidth]{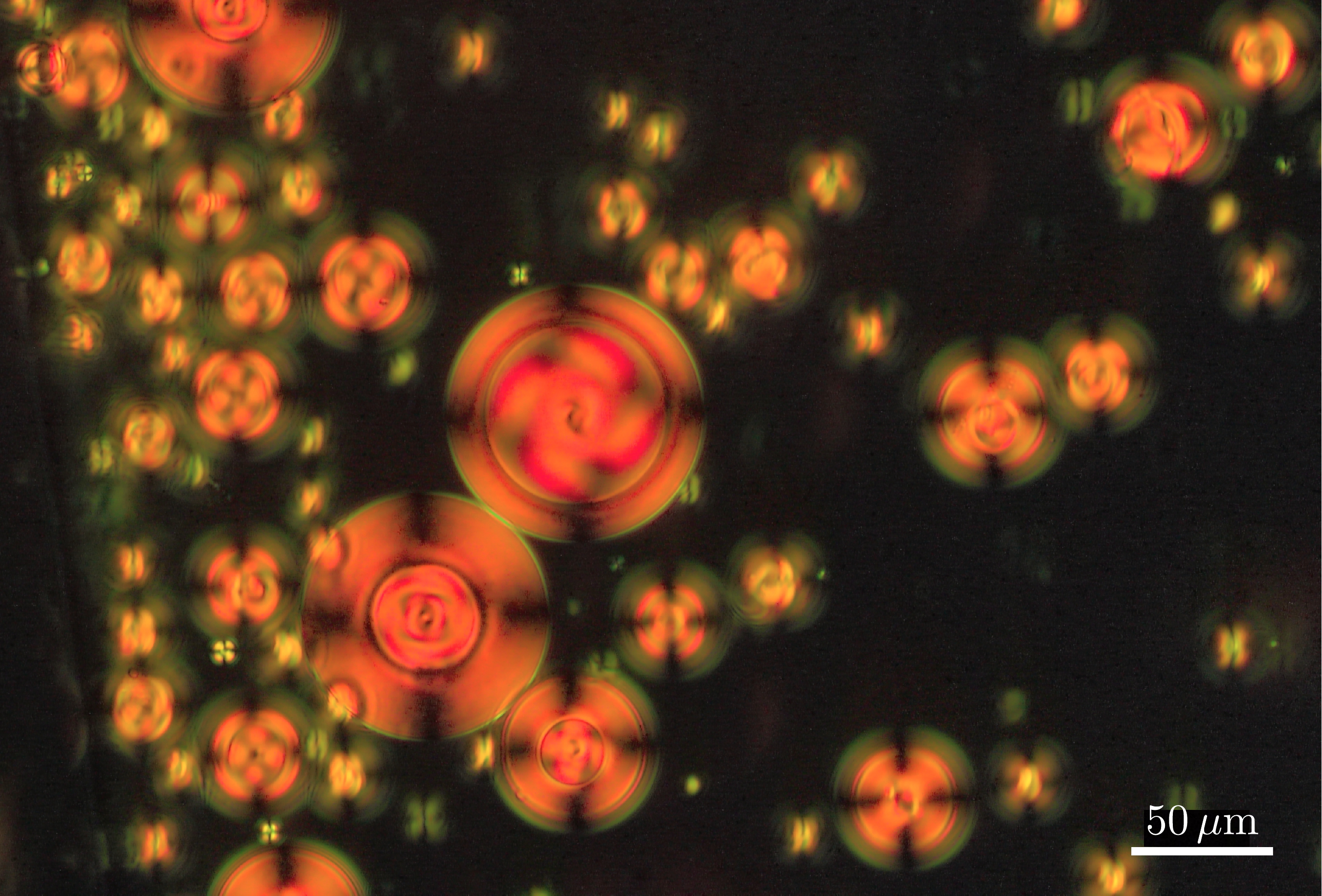}
	\caption{Microcavities enclosing a solution of SSY in water at concentration $c=30\,\mathrm{wt}\%$ and temperature  $\temp=25\,^\circ\mathrm{C}$. }
	\label{fig:cavities}
	\end{subfigure}
	$\qquad\ $
	\begin{subfigure}[c]{0.45\linewidth}
	\centering
	\includegraphics[width=0.4785\linewidth]{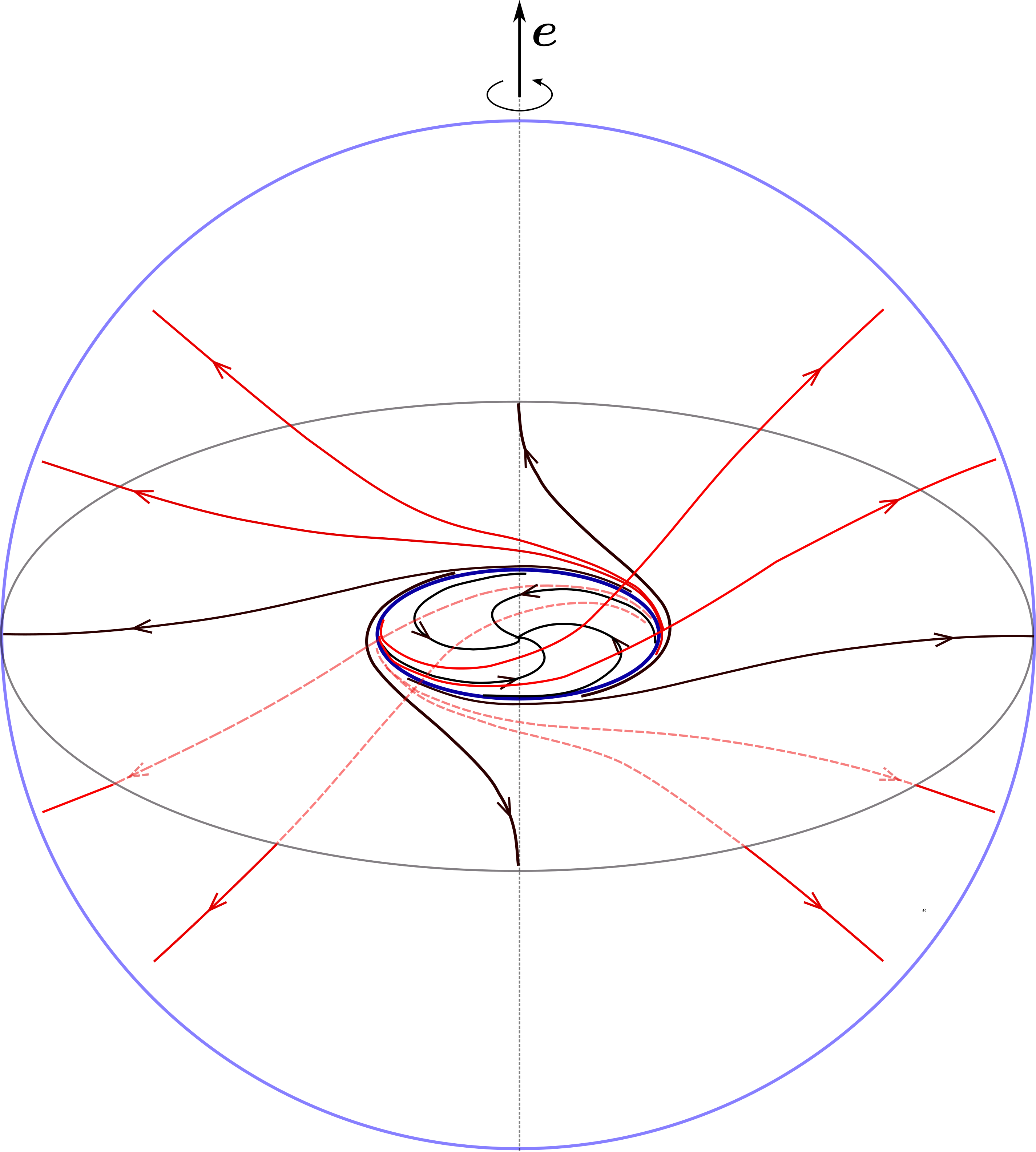}
	\caption{Sketch of the field lines of $\nT$ in \eqref{eq:twisted_hedgehog}. Black lines lie on the equatorial plane, while red lines come out of it.}
	\label{fig:twisted_hedgehog}
\end{subfigure}
\caption{Spherical microcavities enclosing a chromonic solution and enforcing homeotropic boundary conditions. Each cavity allegedly host the same twisted hedgehog, but its symmetry axis is differently oriented relative to the observer. For the cavities viewed along the hedgehog's symmetry axis, the inversion ring is easily identified as a circle; it does not look circular for the cavities viewed askew. In Fig.~\ref{fig:twisted_hedgehog} the inversion ring is depicted in blue; it is present on the equatorial plane (orthogonal to the symmetry axis $\e$). Figs. reprinted by \cite{paparini:spiralling,ciuchi:inversion}.}
\label{fig:tot_splitting}
\end{figure}

The quartic twist theory was applied in \cite{paparini:spiralling,ciuchi:inversion} to describe the \emph{twisted hedgehog} that forms within a spherical cavity enforcing homeotropic alignment on its boundary. It is known since the seminal work of Lavrentovich and Terentiev \cite{lavrentovich:phase} that for a splay constant $K_{11}$ sufficiently larger than the twist constant $K_{22}$, the \emph{radial} hedgehog becomes unstable and acquires a twisted texture (with either sign of chirality equally likely to emerge), see Fig.~\ref{fig:twisted_hedgehog} for one chiral variant of it\footnote{the other is obtained by reversing the sign of $\alpha$ in \eqref{eq:twisted_hedgehog}}. The whole 3D picture is obtained by rotating this drawing about the symmetry axis $\e$.

0.55

More precisely, as proved in \cite{cohen:weak,kinderlehrer:second,rudinger:twist}, the radial hedgehog loses local stability whenever
\begin{equation}
\label{eq:instability_inequality}
K_{11}>K_{22}+\frac18K_{33}.
\end{equation}
The original proof of this inequality was given for the Oseen-Frank elastic theory; however, it also applies to the theory associated with the energy density in  \eqref{eq:quartic_free_energy_density}, as both theories share the same second variation of the energy functional \cite{paparini:stability}. For CLCs, inequality \eqref{eq:instability_inequality} holds because $K_{11}$ and $K_{33}$ are customarily comparable, whereas $K_{22}$ is much smaller (nearly by an order of magnitude).

The field lines shown in Fig.~\ref{fig:twisted_hedgehog} were obtained in \cite{paparini:spiralling} by minimizing the total elastic free energy on the trial family of fields defined as
\begin{equation}
	\label{eq:twisted_hedgehog}
	\nT(\x):=\R(\alpha(r))\frac{\x}{r},
\end{equation}
where $\x$ is the position vector, $r:=|\x|$, and $\R(\alpha)$ denotes the rotation of angle $\alpha$ about a symmetry axis $\e$. A remarkable feature of the twisted hedgehog is the \emph{inversion ring} that the nematic director field $\n$ exhibits on the plane orthogonal to the symmetry axis of the distortion texture. There, the direction of winding of the spiralling field lines of $\n$ changes; it has the optical appearance of a disclination  (see also Fig.~\ref{fig:cavities}), but it is \emph{not} a defect, as  $\n$ is continuous there. The radii of interest are denoted by $r^\ast$ for the inversion ring and 
$R$ for the spherical cavity.

In \cite{ciuchi:inversion} an experimental validation of the quartic theory is presented. This is achieved by extracting measures of the radius of the inversion ring produced in a number of spherical cavities trapping a solution of Sunset yellow (SSY) in water ($c=30, \, 31.5\,\mathrm{wt}\%$ and $\temp=25\,^\circ\mathrm{C}$) inside a polymeric matrix enforcing homeotropic anchoring on the nematic director (see Fig.~\ref{fig:cavities}). Observations of the samples at the optical microscope were made immediately after having prepared the cell, as well as one and two days after. 

\begin{figure}
		\centering
		\includegraphics[width=0.31\linewidth]{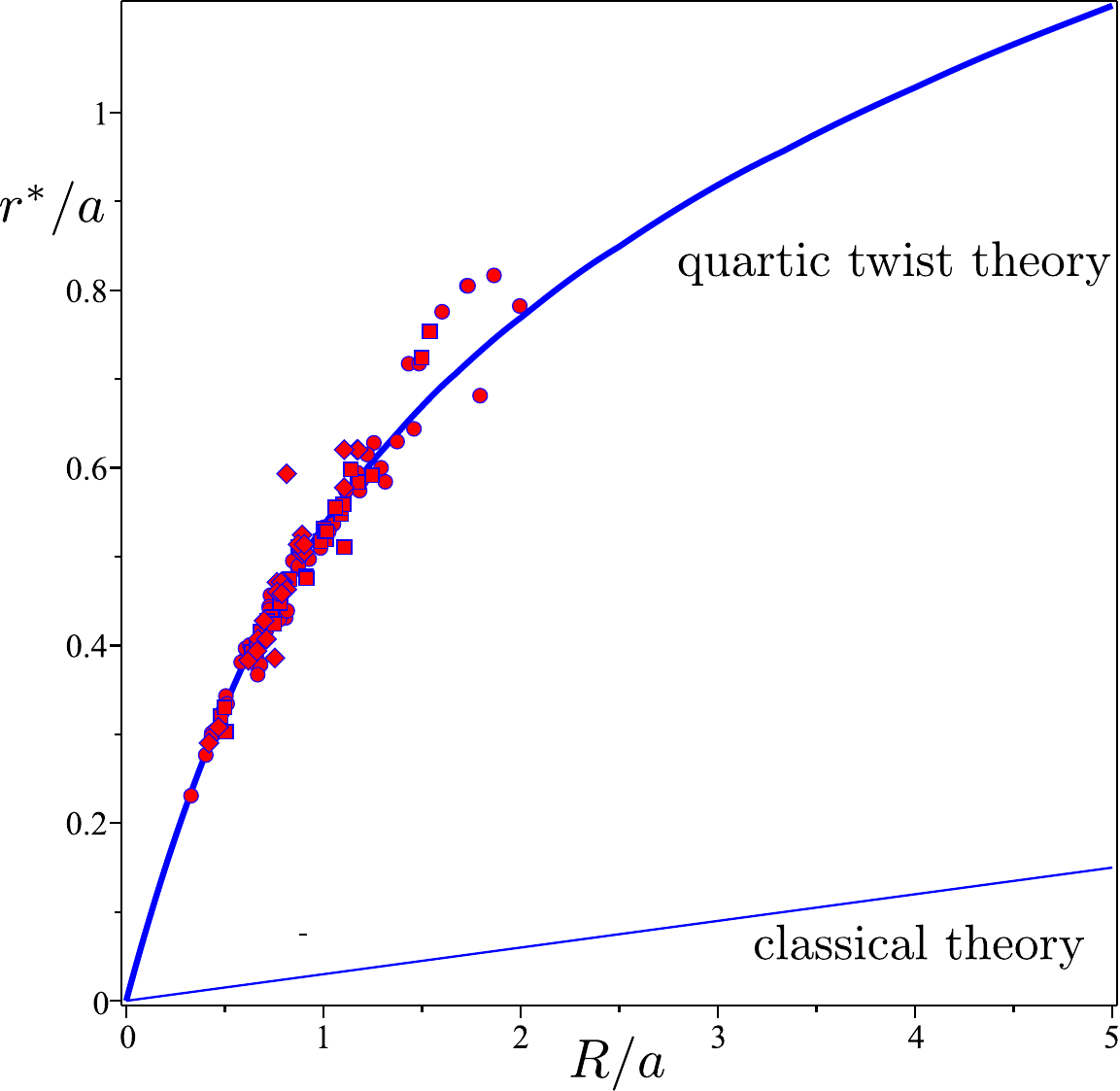}
			\caption{Data collected in all the observations of microcavities containing a SSY solution at concentration $c=30\,\mathrm{wt}\%$  and temperature $\temp=25\,^\circ\mathrm{C}$. Diamonds refer to samples just prepared, boxes to  samples observed after one day,  dots to samples observed after two days. Straight lines apply to the classical Oseen-Frank theory; they are recovered asymptotically by the quartic twist theory in the limit as $R/a\to\infty$. Fig. reprinted by \cite{ciuchi:inversion}.}
	\label{fig:r_star_vs_R}
\end{figure}

From direct measurements of $r^\ast$ and $R$ for each cavity that could be clearly discerned in images like the one in Fig.~\ref{fig:cavities}, they extracted the phenomenological length $a$ featured by the quartic theory. The data collected for $r^\ast$, $R$, and $a$ were then conveniently organized on the universal graph (depending only on $k_1$ and $k_3$ in \eqref{eq:scaled_elastic_constant}) in Fig.~\ref{fig:r_star_vs_R} that plots the ratio $r^\ast/a$ against $R/a$. The quartic theory predicts a monotonic function that approaches $0$ as $R/a\to0$ and grows nonlineraly until becoming asymptotically linear for large $R/a$; it recovers asymptotically the straight line predicted by the classical theory.
It is apparent that for chromonics the Oseen-Frank theory fails to represent the data, whereas the proposed quartic twist theory seems in good agreement with both experiments, more so for smaller cavities than for larger ones, for which the inversion ring tends to be larger than predicted by theory.
Moreover, \cite{ciuchi:inversion} compares the qualitative features of the quartic theory with experiments and finds them in good agreement (see, for example, Fig.~10 in \cite{ciuchi:inversion}).

\subsection{Estimates of $a$}\label{sec:estimates}
The phenomenological length $a$ features in the proposed quartic theory; it is a material parameter, supposedly depending on both temperature and concentration of the chromonic solution. Estimates of $a$ are extracted in \cite{ciuchi:inversion} from direct measurements of the inversion rings in spherical cavities enclosing aqueous solutions of SSY at two different concentrations. These estimates, corresponding to the physical conditions of different observations, are summarized in Table~\ref{tab:a}.
 \begin{table}
\begin{center}
\begin{tabular}{|c|c|c|c| }
\cellcolor{lightgray}{SSY concentration} & \cellcolor{lightgray}{Samples just prepared}  & \cellcolor{lightgray}{Observed after one day} & \cellcolor{lightgray}{Observed after two days}  \\ 
\rule{0pt}{2ex}
$30\,\mathrm{wt}\%$  &  $a\approx 47 \, \mu\mathrm{m} $ &  $a\approx 41 \, \mu\mathrm{m} $ &  $a\approx 33 \, \mu\mathrm{m} $ \\
\hline
$31.5\,\mathrm{wt}\%$ & $a\approx 54 \, \mu\mathrm{m} $ &  $a\approx 46 \, \mu\mathrm{m} $ &  $a\approx 35 \, \mu\mathrm{m}$ \\
 \hline
\end{tabular}
\end{center}
\caption{Estimates of $a$ for SSY solutions in the physical conditions of different observations.}
\label{tab:a}
\end{table}
The values of $a$ are found to exhibit a small dispersion compared to that of radii; this is a testament to the material nature of $a$, which depends only on the physical conditions in which the observation takes place. Two trends emerge clearly from these findings: $a$ increases with concentration and decreases as time elapses since cell preparation.

It is worth noting that for SSY under the same physical conditions as in the first experiment at a concentration of $30\,\mathrm{wt}\%$, a value of $a\approx6.4\,\mu\mathrm{m}$ was estimated from the director field observations of\cite{davidson:chiral} in cylinders with planar degenerate anchoring conditions on their lateral boundary. 
The difference (by nearly one order of magnitude) between this estimate and those summarized in Table~\ref{tab:a} might result from the different experimental settings. By the disparity in the number of data collected here and in \cite{davidson:chiral} and the fact that in \cite{paparini:elastic} two fitting parameters ($a$ and $K_{24}$) are needed to be determined instead of one, the estimates summarized in Table~\ref{tab:a} are likely more reliable.

\section{$2D$ geometries and wide nematic–isotropic coexistence region}\label{sec:2dgeometries}
Recent times have seen a surge of interest in CLCs, mainly because they are soluble in water, and so promise to have valuable applications in life sciences \cite{park:lyotropic}.  Indeed, success has already been granted to the use of CLCs to detect the presence of toxins and cancer biomarkes in simple devices \cite{shiyanovskii:real-time,woolverton:liquid,shaban:label-free}. Determining elastic constants, surface tensions at the nematic-isotropic solution interface, and anchoring strengths for rigid substrates has thus become a priority in the characterization of these materials.

In conventional amphiphile/water systems, the temperature/composition phase diagrams are often complex---and can show a wide 
	range of patterns of aggregation---spherical micelles, cylindrical columns, 
	layered structures complex cubic phases---with the additional factor of 
	there being both oil-in-water and water-in-oil inverse structures. What perhaps distinguishes CLCs is that aggregation starts at very low concentrations and that aggregates are columns, although with variants \cite{lydon:chromonic}.
 Some are stacks of single molecules, others have more than a molecule 
 in their cross-sections~\cite{ogolla:assembly}; it is the variability in size and shape of the supra-molecular columns that makes CLCs so unique.
The aggregation process is \emph{isodesmic}, as the energy gain in adding a unit to a preexisting column 
(typically between $5$ and $10$ $kT$) does not depend on the length of the column. 
The isodesmic nature of the process results in a broad length column distribution, which is prone to the action of temperature. 
When the temperature is increased, the concentration of longer assemblies decreases. This is reflected by the elastic properties
of the phase, in a way that ordinary lyotropics do not exhibit~\cite{ogolla:temperature}. Further increasing the temperature results into a first order nematic-isotropic transition with a wide coexistence region ($5$-$10\,^\circ \mathrm{C}$). Conversely, when the temperature is decreased, short, disordered columns in the isotropic phase tend to grow and aggregate, eventually separating from the parent isotropic solution to form islands of ordered phase.

In this Section, we shall only be concerned with mathematical models provided for chromonic droplets in two space dimensions, inspired by the experimental settings explored by Kim et al.~\cite{kim:morphogenesis}, and Yi et at.~\cite{yi:orientation}. There, bipolar CLC droplets in the nematic phase appeared surrounded by the isotropic phase, and confined within two parallel plates at the distance $h$ from one another.
Substrates like those in \cite{kim:morphogenesis}, for which no preferred orientation is imposed on the bounding plates, are referred to as \emph{degenerate}, whereas substrates like those in \cite{yi:orientation}, which induce one and the same parallel uniform easy axis $\e$ in the nematic phase, are called \emph{aligning} (alignment was achieved by use of topographic patterns that had already been characterized for thermotropic liquid crystals \cite{yoon:organization,kim:alignment,behdani:alignment}).

The experimental settings in \cite{kim:morphogenesis,yi:orientation} suggest a few assumptions to be adopted in the theoretical scene, \cite{paparini:shape,paparini:geometric,kim:morphogenesis,koizumi2023:topological}. First, the director field $\n$ in the nematic phase is parallel to the bounding plates throughout the cell (and so it is a planar field). Second, each nematic region 
will be considered as a cylindrical island $\body$ of prescribed volume $V_0$ occupying the whole gap between the bounding plates. Third, the cross-section $\region$ of $\body$ will be mirror symmetric about two orthogonal
axes, one joining the possible sharp tips of the boundary. For aligning substrates in~\cite{yi:orientation}, this axis coincides with the easy axis $\e$.
In a mathematical language, the island $\body$ is represented as $\region\times[-\frac{h}{2},\frac{h}{2}]$, where the cross-section $\region$ is a region with piecewise smooth boundary $\partial\region$. The isoperimetric constraint \eqref{eq:isoperimetric_constraint} on the volume of $\body$ translates into a constraint on the area of $\region$,
\begin{equation}
\label{eq:area_constraintA0}
A(\region)=A_0,
\end{equation}
where $A$ is the area measure and $A_0=V_0/h$.

It has to be noted that in this two-dimensional context we do not delve on the possibly controversial issue discussed in Sec.~\ref{sec:conundrum} concerning the paradoxical consequences of violating \eqref{eq:Ericksen_inequalities_2} for the equilibrium shape of CLC droplets surrounded in three space dimensions by an isotropic fluid. Here, since the director field $\n$ is planar, it lies everywhere parallel to a given plane and is independent of the coordinate orthogonal to that plane. For such a field (see also \cite{pedrini:relieving}),
\begin{equation}
	\label{eq:characteristics_planar}
	T=0\quad\text{and}\quad S^2=4q^2,
\end{equation}  
and so \emph{both} $\WOF$ and $\WQT$ reduce to 
\begin{equation}
	\label{eq:planar_reduced_energy}
	W=\frac12K_{11}S^2+\frac12K_{33}B^2,
\end{equation} 
which is a well-behaved positive-definite energy density for
\begin{equation}
	\label{eq:reduced_inequalities}
	K_{11}>0\quad\text{and}\quad K_{33}>0,
\end{equation}
the only inequalities needed below. Therefore, the bulk free energy stored in the cylindrical island $\body$ results to be given by the functional
\begin{equation}
\label{eq:free_energy_2d}
\free_\mathrm{b}[\region;\n]:=\frac{h}{2}\int_{\region}[K_{11}(\diver\n)^2+K_{33}|\n\times\curl\n|^2]\dd A,
\end{equation}
where $\dd A$ is the area element. At the interface between $\body$ and the surrounding isotropic solution a surface energy arises, represented by the 
classical Rapini-Papoular formula \cite{rapini:distortion} as in the three-dimensional case (ref. to eq. \eqref{eq:surface_energy}). In the two-dimensional context it is given by
\begin{equation}
	\label{eq:Rapini_Papoular_formula}
	\free_{\mathrm{s}}\left[\region;\n\right]=h\int_{\partial\region}\gamma[1+\omega(\n\cdot\normal)^2]\dd\ell,
\end{equation}
where $\normal$ is here the outer unit normal to the interface $\partial\region$, and $\dd \ell$ is the length element. Here, $\omega$ is assumed to be positive, so that the interfacial energy density (per unit area) is minimized when $\n$ is tangent to the interface. Where a CLC is in contact with an aligning substrate, as in \cite{yi:orientation}, the anchoring energy $\free_{\mathrm{a}}$ is represented similarly to \eqref{eq:Rapini_Papoular_formula} as
\begin{equation}
	\label{eq:anchoring_energy}
	\free_{\mathrm{a}}\left[\region;\n\right]=\int_{\region}\sigma_0[1-(\n\cdot\e)^2]\dd A
\end{equation}
where $\sigma_0>0$ is the \emph{anchoring strength} and $\e$ is a unit vector designating the easy axis. In \eqref{eq:anchoring_energy}, the anchoring energy density is normalized so as to have zero minimum. 
The total free-energy functional $\free$ which describes a chromonic island $\body$ with cross-section $\region$ is then obtain by adding all energy contributions discussed above,
\begin{equation}
\label{eq:free_energy_functional_aligning}
\free[\region;\n]:=\free_{\mathrm{b}}\left[\region;\n\right]+\free_{\mathrm{a}}\left[\region;\n\right]+\free_{\mathrm{a}}\left[\region;\n\right].
\end{equation}
Note that for degenerate substrates, in light of the isoperimetric constraint \eqref{eq:area_constraintA0} which prescribes the area $A_0$ of the admissible domains $\region$, the additional anchoring energy can be treated as an inessential additive constant.

\subsection{Morphogenesis of defects and topological shape transformation in CLCs}\label{sec:morphogenesis}
Morphogenesis in living systems involves topological shape transformations that are highly unusual in the inanimate world. 
The interplay between surface, anchoring and bulk forces in changing topology is far from being understood. Liquid crystal droplets provide a simple yet insightful system, where the effect of such forces on both shape and internal structure can, in principle, be tractable, \cite{koizumi2023:topological} 

This Section focuses on two key contributions addressing the morphogenesis of nuclei and topological defects during phase transitions in CLCs \cite{kim:morphogenesis} and topological shape transformation caused by increasing the splay-to-bend ratio $K_{11}/K_{33}$ \cite{koizumi2023:topological}. Throughout this section, the substrates are assumed to be degenerate, in accordance with their experimental setups.

\begin{figure}
	\centering
	\begin{subfigure}[c]{0.45\linewidth}
		\centering
		\includegraphics[width=0.9\linewidth]{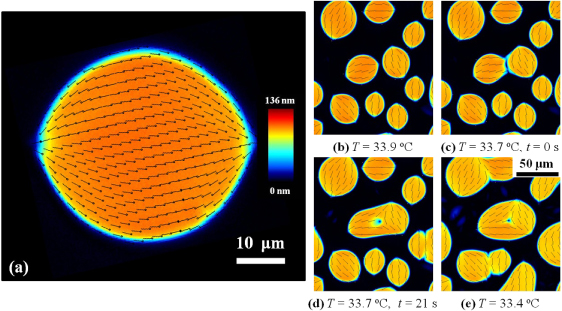}
		\caption{PolScope textures of nucleating N tactoids during the I-to-N phase transition in DSCG. Fig. reprinted by \cite{kim:morphogenesis}.}
	\label{fig:kim_tactoid}
	\end{subfigure}
	$\qquad\ $
	\begin{subfigure}[c]{0.48\linewidth}
		\centering
		\includegraphics[width=\linewidth]{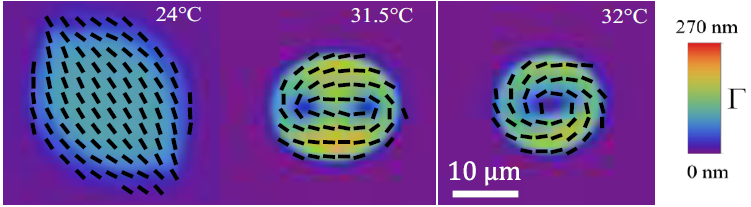}
		\caption{PolScope textures of temperature-triggered tactoid-to-toroid transformation. Fig. reprinted by \cite{koizumi2023:topological}.}
		\label{fig:lavrentovich_tactoid}
	\end{subfigure}
	\caption{Examples of morphogenesis of topological defects and topological shape transformation in CLCs.}
	\label{fig:morphogenesisofdefects}
\end{figure}

In \cite{kim:morphogenesis}, the authors
illustrate how the balance of anisotropic surface energy and bulk elasticity can shape complex morphogenetic developments of tactoids and topological defects emerging in phase transitions between the isotropic (I) and nematic (N) phases of CLCs. The observed point defects in $2D$ are of two types (see \cite{kim:morphogenesis} for their topological description); point defect disclinations of integer and semi-integer strength are located in the interior of the N domain, whereas point defects-boojums of continuously defined topological charges are located at the cusps, which are points separating two differently tilted
shoulders of the I-N interface.
The reason for the existence of defects is the anisotropic nature of the I-N interface that favours $\n$ being tangential, as is clear from the image of the large tactoid in Fig.~\ref{fig:kim_tactoid}(a).
As the temperature is lowered, the tactoids grow and coalesce. Most of the time, the tactoids coalesce in pairs and the new tactoids eventually regain the same shape. However, the merger can also produce nontrivial topological defects in the bulk, similarly to the Kibble mechanism. For
example, Figs.~\ref{fig:kim_tactoid}(c)–(e) show the formation of an $m=-1/2$ interior disclination resulting from the coalescence of two nearly orthogonally aligned tactoids, producing a three-cusp tactoid with boojums of positive strength $0<m<1/2$. Coalescence may also form \emph{negative} cusps of protruding I phase, carrying boojums in the N part of negative charge $-1/2<m<0$, as clearly visible in Fig.~\ref{fig:kim_tactoid}(e).

The presence of interior disclinations influences the number of cusps featured by the closed I–N interface. In \cite{kim:morphogenesis}, conservation laws are derived that relate the excess number of positive and negative cusps, $c^+$ and $c^-$, to the topological strengths $m_j$ of the interior disclinations, with $j=1,\dots,n$. The relation is
\begin{equation}
\label{eq:conservationlaw}
c^+-c^-=2\left(1-\sum_{j=1}^n m_{j}\right).
\end{equation}
Additional conservation laws for defects associated with simply connected isotropic tactoids are also given in \cite{kim:morphogenesis}, but fall outside the scope of this review. 

The experimental explorations in \cite{kim:morphogenesis} brought into evidence that the core of disclinations in CLCs might be very large and extend over macroscopic length scales accessible for optical characterization. Taking advantage of this fact, \cite{zhou2017:fine} explores the fine structure of the disclination cores at both the micron and sub-micrometer scales through optical and electron microscopy; the director $\n$ and the scalar order parameter $s$ associated with the degree of orientational order show a profound
change in the core region. This unexpected core structure is explained by a strong coupling between
the gradients of $\n$ and $s$ in the free energy.
The reader is also referred to \cite{zhang2018:computational} for simulations of the topological defects produced by tactoids at domain junctions as in Fig.~\ref{fig:kim_tactoid}, employing a model in which the degree of order $s$, the director $\n$, and the interfacial normal $\normal$ serve as state descriptors.

In \cite{koizumi2023:topological} (see also \cite{lavrentovich2024:splaybend}) it is illustrated that CLC nematic droplets coexisting with the isotropic phase change their shape from a simply-connected tactoid as that in Fig.~\ref{fig:kim_tactoid}(a) to a topologically distinct toroid of radius $a$ as a result of temperature or concentration variation, Fig.~\ref{fig:lavrentovich_tactoid}. The transformation is driven by an increase of the splay-to-bend ratio $K_{11}/K_{33}$. The two shapes are topologically distinct, as described by Euler characteristic $\chi$, calculated as $\chi=2-2g$, where $g$ is the number of “handles”; a sphere has no handles, thus $\chi=2$, while a torus is a single handle, thus $\chi=0$. That transformation starts with the detachment of the two surface point defects-boojums from the cusps of tactoid, making them two disclinations of strength $+1/2$ each. The disclinations approach each other and coalesce, forming a toroid with a large central isotropic region. By \eqref{eq:area_constraintA0}, the surface area of the tactoid equals that of the torus, and so $A_0=\pi a^2$.

As follows from \eqref{eq:free_energy_2d}, when $K_{11}$ is similar to the bend modulus $K_{33}$, the
droplet accommodates both splay and bend of the director $\n$ within a simply connected tactoid, whose bulk free-energy $\free_{\mathrm{b}}$ is estimated to be
\begin{equation}
\label{eq:free_bulk_tactoidlavrentovich}
\free_{\mathrm{b}}^{\mathrm{tac}}\approx\frac{\pi h}{2}K_{11}\ln\left(\frac{2a}{r_{\mathrm{c}}}\right)+\frac{\pi h}{2}K_{33}(1-\ln 2),
\end{equation}
where $r_{\mathrm{c}}$ is the radius of the core of the boojums. 
When $K_{11}$ increases, the droplet could afford only bend, resulting in a torus-like shape with a hole in the center and with energy
\begin{equation}
\label{eq:free_bulk_toruslavrentovich}
\free_{\mathrm{b}}^{\mathrm{tor}}=\pi hK_{33}\ln\left(\frac{a}{r_\mathrm{i}}\right),
\end{equation}
where $r_\mathrm{i}$ is the radius of the isotropic core. Surface tension also contributes to the transformation scenario, though less significantly than elasticity. Numerical simulations in \cite{koizumi2023:topological}, which account for both elastic and surface effects, provide a more accurate description of the transition, including the emergence and coalescence of the $1/2$-disclinations.

\subsection{Shape Bistability in $2D$ CLC droplets}\label{sec:shape}
\begin{figure}
\centering
	\begin{subfigure}[c]{0.48\linewidth}
		\centering
		\includegraphics[width=0.65\linewidth]{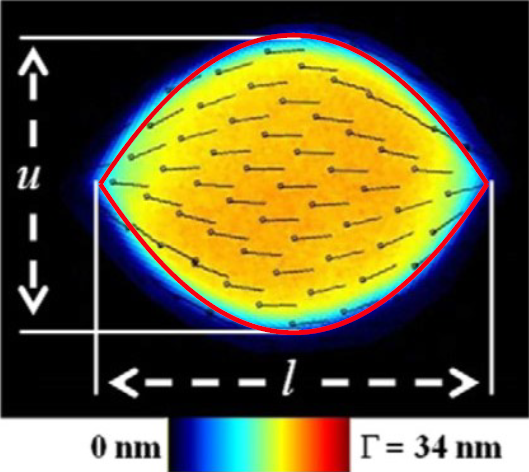}
		\caption{The droplet's equilibrium shape (in red) is contrasted against the shape observed in \cite{kim:morphogenesis}, see Figs.~\ref{fig:kim_tactoid}(a)-(b). Fig. reprinted by \cite{paparini:shape}.}
	\label{fig:Comparison_Kim}
	\end{subfigure}
	$\quad\ $
	\begin{subfigure}[c]{0.48\linewidth}
		\centering
		\includegraphics[width=0.73\linewidth]{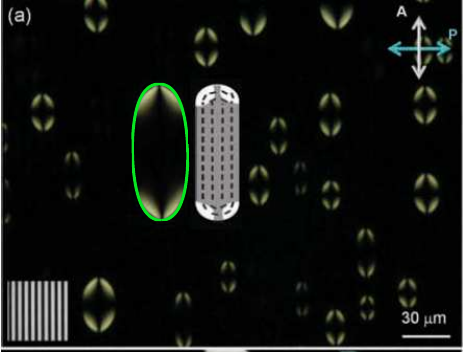}
	\caption{The droplet outlined in green corresponds to the minimizer of the free energy, and is superimposed to the experimental image in \cite{yi:orientation}. Fig. reprinted by \cite{paparini:geometric}.}
	\label{fig:Clark_experiments}
	\end{subfigure}
	\caption{Observations of two-dimensional drops bearing a bipolar director field $\n$ with only in-plane components and point defects of $\n$ at the poles. In the case shown in Fig.~\ref{fig:Comparison_Kim}, the substrates are degenerate, whereas in Fig.~\ref{fig:Clark_experiments} they are aligning. In particular, the white stripes on the bottom left corner of Fig.~\ref{fig:Clark_experiments} designate the orientation of the aligning channels on both bounding substrates.}
	\label{fig:shapesobserved}
\end{figure}

Experiments reported in \cite{kim:morphogenesis} and \cite{yi:orientation} revealed essentially \emph{two-dimensional} droplets bearing a \emph{bipolar} director field $\n$ with only in-plane components and point defects of $\n$ at the poles, see Fig.~\ref{fig:shapesobserved}. In \cite{kim:morphogenesis}, the droplets exhibited the characteristic spindle-like shapes known as tactoids, whereas in \cite{yi:orientation} they predominantly appeared as elongated rods with rounded ends, referred to as \emph{bâtonnets} \cite{paparini:geometric}.

\begin{figure}[h]
	\centering
	\begin{subfigure}[b]{0.2\linewidth}
		\centering
		\includegraphics[width=0.8\linewidth]{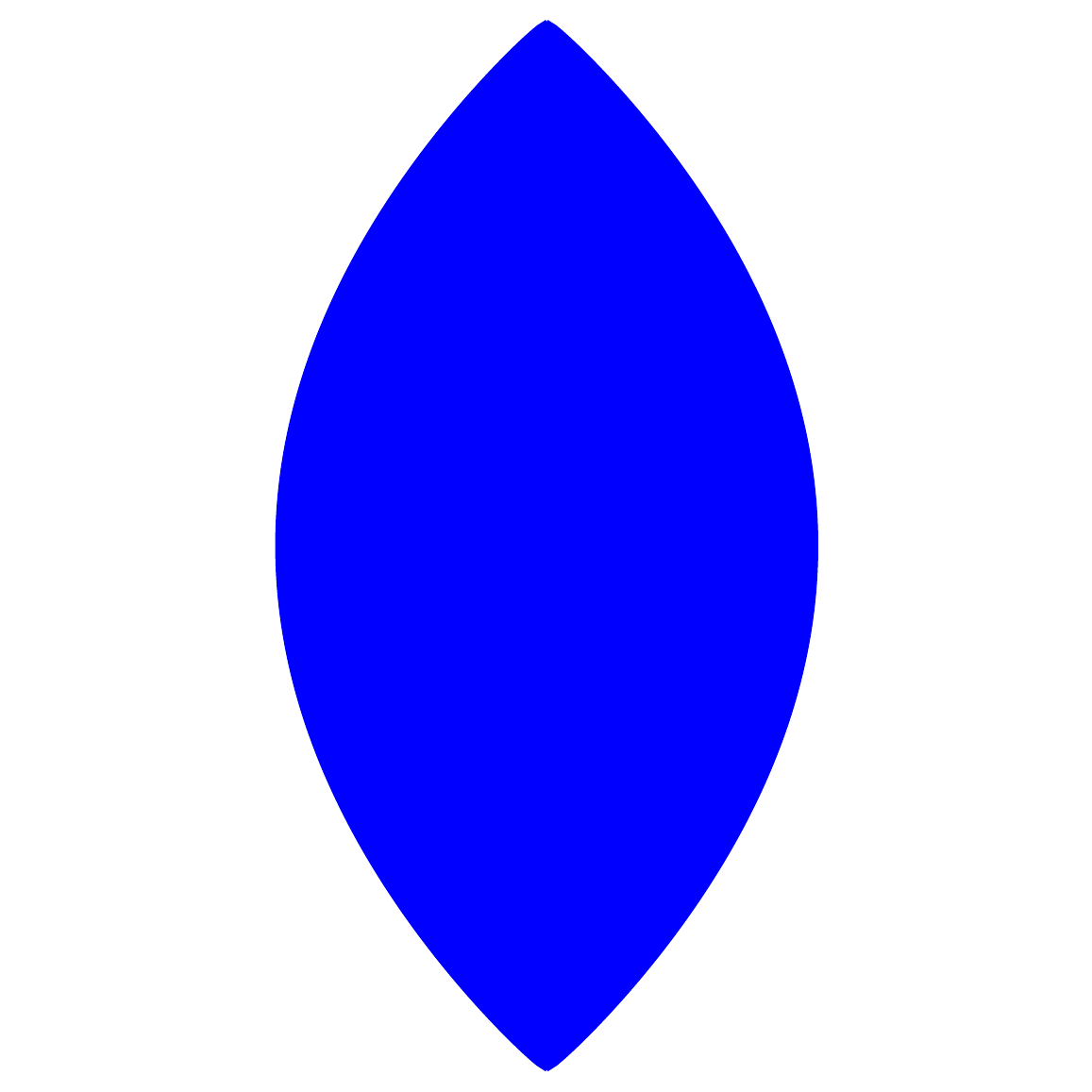}
		\caption{tactoid.}
		\label{fig:tacoid}
	\end{subfigure}
	\quad
	\begin{subfigure}[b]{0.2\linewidth}
		\centering
		\includegraphics[width=0.584\linewidth]{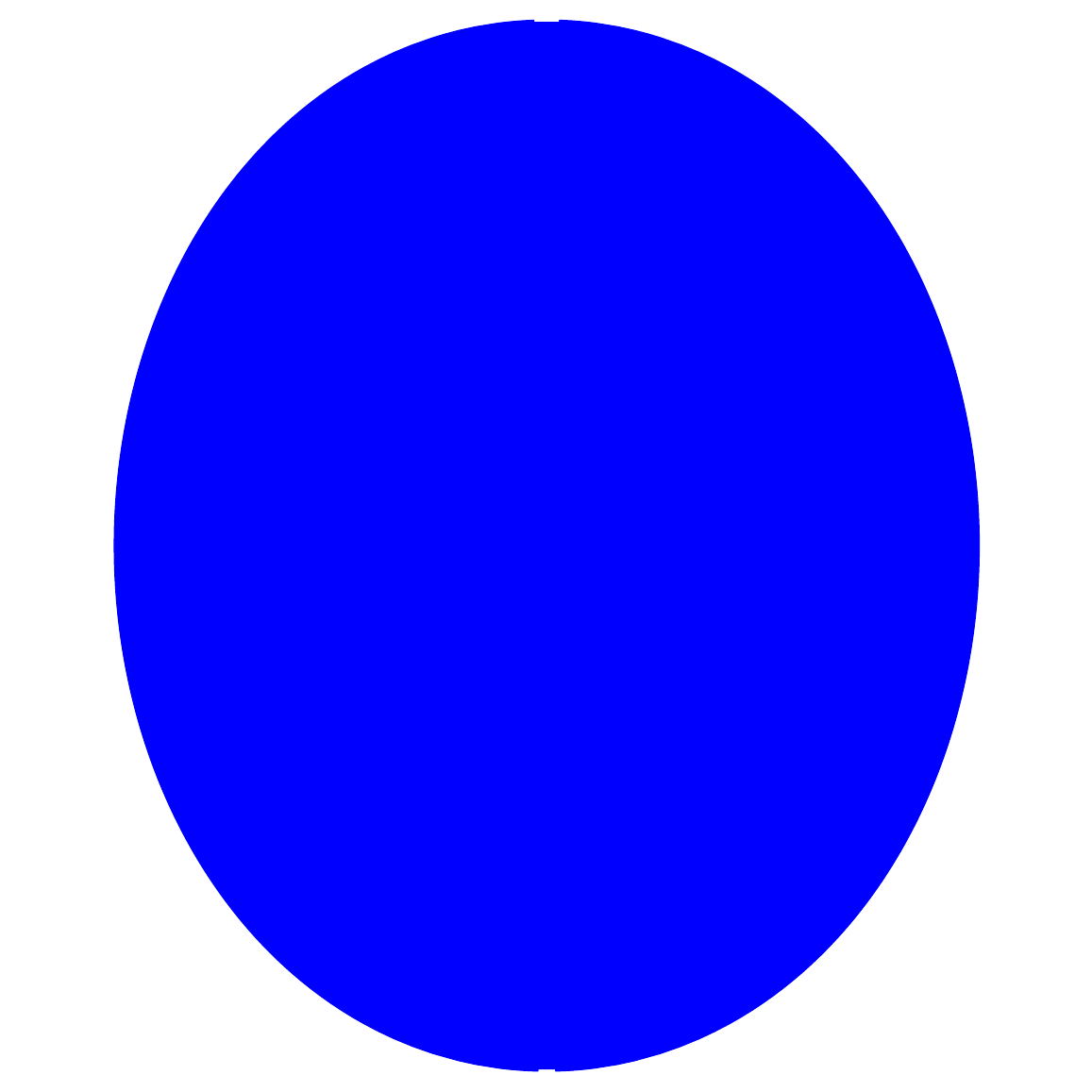}
		\caption{discoid.}
		\label{fig:discoid}
	\end{subfigure}
	\quad
	\begin{subfigure}[b]{0.2\linewidth}
		\centering
		\includegraphics[width=0.8\linewidth]{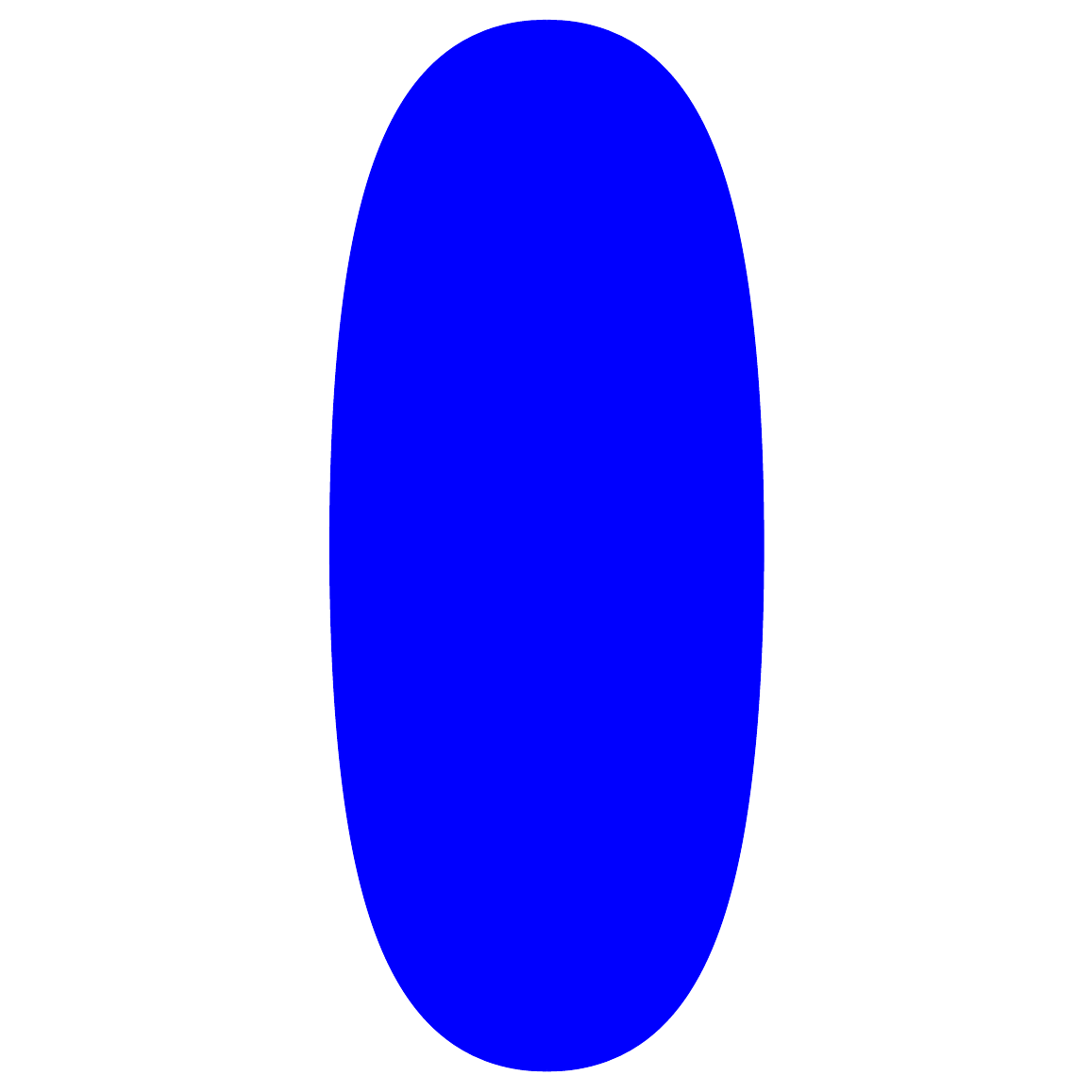}
		\caption{b\^{a}tonnet.}
		\label{fig:batonnet}
	\end{subfigure}
	\caption{Gallery of exemplary shapes, each showing a different type of possible convex minimizer of the free energy. They all have the same area $A_0$. Fig. reprinted by \cite{paparini:geometric}.}
	\label{fig:gallery_aligning}
\end{figure}

In \cite{paparini:shape,paparini:geometric}, the authors extended the study in \cite{kim:morphogenesis} through further theoretical modeling. The analysis revealed that, upon increasing the droplet's area, the equilibrium shape transitions from tactoidal to discoid (smooth), while concave shapes cannot occur at equilibrium. Bâtonnets, on the other hand, are observed only for aligning substrates \cite{paparini:geometric}. Moreover, for both degenerate and aligning substrates, the models predict a regime of \emph{shape coexistence}, where a bipolar tactoid and a bipolar discoid are both local minima of the free energy, with the global minimum shifting from one to the other at a critical value of the droplet's area, where perfect bistability is established. The critical values of the droplet's area that delimit the corresponding shape hysteresis depend on the ratio of the elastic constants  $K_{11}/K_{33}$ of the material, and, for aligning substrates, also on the anchoring strength. 
Representative shapes of the admissible ones are shown in Fig.~\ref{fig:gallery_aligning}.
They are all convex. 

The regime of shape bistability manifests itself for droplets larger than those reported in \cite{kim:morphogenesis} and smaller than those observed in \cite{yi:orientation}; it does not seem to have been observed, at least in \cite{kim:morphogenesis,yi:orientation}. This shape bistability was not predicted for fully three-dimensional droplets \cite{paparini:nematic} and it might be a signature of two-dimensionality.

\subsection{Estimate of the surface tension at the nematic/isotropic interface}\label{sec:estimategamma}
Different, discordant estimates have been given in the literature for the order of magnitude of $\gamma$. For example, in \cite{mushenheim:using} they estimate $\gamma\sim1\,\mu\mathrm{N/m}$, whereas  in \cite{kim:morphogenesis} they give $\gamma\sim10^{2}\,\mu\mathrm{N/m}$, an estimate obtained by applying the pendant drop technique \cite{chen:interfacial,chen:interfacial_1999}. In \cite{paparini:shape}, the very detailed data in \cite{kim:morphogenesis} of two-dimensional droplets sandwiched between degenerate substrates are used to compare the observed shapes with those predicted by their mathematical model, see Fig.~\ref{fig:Comparison_Kim}. Encouraged by their agreement, the authors extracted the following estimate for the surface tension at the nematic/isotropic interface of an aqueous DSCG solution at $16\,\mathrm{wt\%}$,
\begin{equation}
\label{eq:gamma_Kim}
 \gamma\approx 8.9\,\mu\mathrm{N/m},
\end{equation}
which turns out to comparable in order of magnitude to the typical values measured for standard thermotropic liquid crystals ($\sim 10\, \mu\mathrm{N/m}$, see \cite[p.\,495]{kleman:soft})\footnote{The value in \eqref{eq:gamma_Kim} is closer to that found in \cite{faetti:anchoring} at the nematic-isotropic interface of 5CB (see also \cite{faetti:measurements,faetti:nematic}).}.

\subsection{Estimate of CLCs' planar anchoring strength}\label{sec:estimatesigma}
Some estimates of the anchoring strength $\sigma_0$ for chromonics in contact with different substrates are already known: they range from $\sigma_0\sim10^{-1}\,\mu\mathrm{J/m^2}$, for both scratched glasses \cite{mcguire:orthogonal} and rubbed polyimide surfaces \cite{collings:anchoring}, to $\sigma_0\sim10^2\,\mu\mathrm{J/m^2}$, for surfaces lithographed by secondary sputtering \cite{kim:macroscopic}.\footnote{For thermotropic liquid crystals, the  strength of planar anchoring ranges from about $1\,\mu\mathrm{J/m^2}$ to one or two orders of magnitude higher, as shown, for example, in Table~3.1 of \cite{blinov:electrooptic}. On the strongest side is the measurement of \cite{faetti:strong}, based on an improved reflectometric method introduced in \cite{faetti:improved}; for 5CB, it was found that $\sigma_0\sim10^2\mathrm{J/m^2}$.}

In \cite{paparini:geometric}, a method is proposed to determine the planar anchoring strength $\sigma_0$ of a chromonic liquid crystal on a rigid substrate; its distinctive feature is  \emph{geometric}, as it is based on recognition and fitting of the stable equilibrium shapes of droplets surrounded by the isotropic phase in a thin cell with plates enforcing parallel alignments of the nematic director. In particular, the author focus on the droplet shown in Fig.~12a of \cite{yi:orientation} (and highlighted in Fig.~\ref{fig:Clark_experiments}).
Prior knowledge of the surface tension $\gamma$ of the nematic phase at the isotropic interface is required, which can be gained by the study in \cite{paparini:shape} of cells with substrates enforcing planar degenerate anchoring. Accordingly, by using the estimate of $\gamma\approx 8.9\,\mu\mathrm{J/m^2}$ in \eqref{eq:gamma_Kim} for the surface tension of an aqueous DSCG solution (at $c=16\,\mathrm{wt}\%$ and $T=32.5^\circ \mathrm{C}$ \cite{paparini:shape}), they arrive at  the following estimate of the anchoring strength,
\begin{equation}
\label{eq:sigma_Clark}
\sigma_0\approx22\,\mu\mathrm{J/m^2},
\end{equation}
which turns out to be of the same order of magnitude as $\gamma$ for DSCG and intermediate between values  measured with other methods  for the same material. Such an estimate for $\sigma_0$ may be affected by the chosen value of $\gamma$ in \eqref{eq:gamma_Kim}, that is obtained in \cite{paparini:shape} for a DSCG solution at different concentration and temperature than the one studied in \cite{yi:orientation}, but better data are lacking. 

One method employed to measure $\sigma_0$ for chromonics, though not in two-dimensional geometries, uses twist cells with plates promoting planar easy axes at right angles to one another. Within the classical Oseen-Frank theory, measuring the total \emph{twist angle} $\Omega$ across the cell (and how it differs from $90^\circ$) determines $\sigma_0$, once the twist constant $K_{22}$ is known \cite{mcginn:planar}. This method relies on the subtle theory (put forward  by McIntyre~\cite{mcintyre:light,mcintyre:transmission}) relating (in closed form) $\Omega$ to the maximum and minimum transmitted intensity of light with normal incidence propagating (between crossed polarizers) through the cell. 
For example, this method has been applied in \cite{collings:anchoring} and \cite{peng:patterning}. In \cite{collings:anchoring}, the value of $\sigma_0$ at the plates of a cell filled with a DSCG aqueous solution was determined to be less than $1\,\mu\mathrm{J}$, whereas \cite{peng:patterning} reported a larger value of $\sigma_0$, obtained adopting a different substrate. Instead, in \cite{paparini:what}, the experiments performed in \cite{collings:anchoring,peng:patterning} has been re-examined in the light of the quartic twist theory (ref. to Sec.~\ref{sec:quartic}). Also this theory predicts a linear twist between the cell’s plates with an offset angle related to both the anchoring strength $\sigma_0$ and the phenomenological length $a$ featuring in the theory. The data presented in both \cite{collings:anchoring,peng:patterning} turn out to be better fitted by this theory compared to the classical quadratic
one (with a slightly less error); as a result, both $\sigma_0$ and $a$ are determined, the 
former with a value larger than the one found in \cite{collings:anchoring} and much closer to that found in \cite{peng:patterning}. This approach used to measure $\sigma_0$ might therefore serve as an alternative, independent method to rely upon.

\section{Open questions and Future Avenues of Research}\label{sec:conclusion}

The studies reported in this Review suggest several avenues for future investigation, both theoretical and experimental.
From the theoretical side, the following challenges could be addressed.
\begin{itemize}
\item[(T1)] As recalled in Sec.~\ref{sec:stability}, degenerate planar boundary conditions play a key role in stabilizing the ground state of chromonics within the classical elastic theory of Oseen and Frank. This poses the following question: what is the most general class of anchoring conditions capable in rigid containers of preventing the energy from diverging to negative infinity?
\item[(T2)] The quartic twist theory proposed in Sec.~\ref{sec:quartic} introduces a phenomenological length $a$, whose microscopic origin is as problematic as the
origin of spontaneous (single) twist in cholesteric liquid crystals. A demanding question is how $a$ is related to the structure of the microscopic
constituents of the material. 
In Table~\ref{tab:a}, the average of $a$ for each observation is found to increase with concentration and decrease with the time elapsed after sample preparation. 
One may interpret this result as suggesting that molecular aggregates become shorter on average (hence larger in number) as the time progresses. 
A theoretical validation of this conjecture could be found by building 
a kinetic model for chromonic molecular aggregation, possibly along lines similar to those that started to be traced in \cite{pergamenshchik:kinetic,pergamenshchik:statistical}. 
\end{itemize}

On the other side, the outcomes generated by the models discussed in this review can serve to propel future experiments on CLCs.
\begin{itemize}
\item[(E1)] CLC droplets can undergo a topological shape change from sphere-like tactoids to tori driven by elastic anisotropy. This mechanism may shed light on topological transformations in morphogenesis and inspire new strategies for shaping soft materials.
\item[(E2)] The data currently available in the literature do not cover the range of predicted bistability for two-dimensional droplets between either degenerate or aligning substrates. For both cases, models can estimate the critical area that a droplet should reach to display an abrupt transition from tactoid to discoid, and viceversa.  It remains to be seen whether a controlled growth (or decrease) in the droplet’s size can be realized to observe neatly this transition.
\item[(E3)] Methods to determine the isotropic surface tension at the nematic–isotropic interface $\gamma$ and the chromonics' anchoring strength $\sigma_0$, based on extensive experimental studies of the shape of two-dimensional bipolar CLC droplets confined between two parallel plates, have been presented.  These estimates seem to promise more accuracy than the rough evaluation of order of magnitude available in the literature for this material and its chromonic siblings. The theories proposed could be used for a systematic determination of $\gamma$ and $\sigma_0$ for different temperatures and concentrations.
\end{itemize}

Although not addressed in this review, the studies in \cite{paparini:singulartwist,paparini:singulardamped} reveal that the quartic structure of the elastic energy (ref. to Sec.~\ref{sec:quartic}) is responsible for the formation of shock waves in a twist director mode. Such a wave breaking could serve as an experimental signature of the validity of the elastic quartic twist theory.

\begin{acknowledgments}
I am really grateful to Prof. Ingo Dierking for having invited this review.\\
I am member of the Italian \emph{Gruppo Nazionale per la Fisica Matematica} (GNFM), which is part of INdAM, the Italian National Institute for Advanced Mathematics. I gratefully acknowledge partial financial support provided for this work by GNFM.
\end{acknowledgments}

\bibliography{Chromonics}

\end{document}